\documentclass[iop]{emulateapj}
\usepackage{natbib}
\usepackage{xcolor}

\shorttitle{The Binary System 3~Pup} \shortauthors{A.~Miroshnichenko, S.~Danford, S.~Zharikov,  et~al.}

\begin{document}

\title{Properties of Galactic B[e] Supergiants: V. 3\,Pup -- constraining the orbital parameters and modeling the circumstellar environments.}

\author{A.~S.~Miroshnichenko$^{1,2,3}$}
\affil{$^1$Department of Physics and Astronomy, University of North Carolina at Greensboro, P.O. Box 26170, Greensboro, NC 27402--6170, USA}
\affil{$^2$Pulkovo Astronomical Observatory of the Russian Academy of Sciences, Pulkovskoe shosse 65-1, St. Petersburg, 196140, Russia}
\affil{$^3$Fesenkov Astrophysical Institute, Observatory, 23, Almaty, 050020, Kazakhstan}

\author{S.~Danford$^1$}
\affil{$^4$Department of Physics and Astronomy, University of North Carolina at Greensboro, P.O. Box 26170, Greensboro, NC 27402--6170, USA}

\author{S.~V.~Zharikov$^4$}
\affil{$^4$Instituto de Astronomia, Universidad Nacional Aut\'onoma de M\'exico, Apdo. Postal 877, Ensenada, 22800, Baja California, M\'exico}

\author{V.~G.~Klochkova$^5$, E.~L.~Chentsov$^5$}
\affil{$^5$Special Astrophysical Observatory of the Russian Academy of Sciences, Nizhnyj Arkhyz, Zelenchukskiy region,
Karachai-Cherkessian Republic, 369167, Russia}

\author{D.~Vanbeveren$^6$}
\affil{$^6$Astronomy and Astrophysics Research Group, Vrije Universiteit Brussel, Pleinlaan 2, 1050 Brussels, Belgium}

\author{O.~V.~Zakhozhay$^{7}$}
\affil{$^7$Main Astronomical Observatory, National Academy of Sciences of Ukraine, Kyiv 03680, Ukraine}

\author{N.~Manset$^8$}
\affil{$^8$CFHT Corporation, Kamuela, HI 96743, USA}

\author{M.~A.~Pogodin$^2$}
\affil{$^2$Pulkovo Astronomical Observatory of the Russian Academy of Sciences, Pulkovskoe shosse 65-1, St. Petersburg, 196140, Russia}

\author{C.~T.~Omarov$^{3,9}$}
\affil{$^3$Fesenkov Astrophysical Institute, Observatory, 23, Almaty, 050020, Kazakhstan}
\affil{$^{9}$National Center of Space Research and Technology, Shevchenko St. 15, Almaty, 050010, Kazakhstan}

\author{A.~K.~Kuratova$^{10}$, S.~A.~Khokhlov$^{10}$}
\affil{$^{10}$Al-Farabi Kazakh National University, Al-Farabi Ave. 71, 050040, Almaty, Kazakhstan}

\vspace*{1.0cm}
\begin{abstract}
We report the results of a long-term spectroscopic monitoring of the A--type supergiant with the B[e] phenomenon 3\,Pup = HD\,62623.  We confirm earlier findings that it is a binary system. The orbital parameters were derived using cross-correlation of the spectra in a range of 4460--4632 \AA, which contains over 30 absorption lines. The orbit was found circular with a period of $137.4\pm0.1$ days, radial velocity semi-amplitude $K_{1} = 5.0\pm0.8$ km\,s$^{-1}$, systemic radial velocity $\gamma = +26.4\pm2.0$ km\,s$^{-1}$, and the mass function $f(m) =  (1.81^{+0.97}_{-0.76})\times10^{-3}$ M$_{\odot}$. The object may have evolved from a pair with initial masses of $\sim$6.0 M$_{\odot}$ and $\sim$3.6 M$_{\odot}$ with an initial orbital period of $\sim$5 days. Based on the fundamental parameters of the A-supergiant (luminosity $\log$ L/L$_{\odot} = 4.1\pm$0.1 and effective temperature T$_{\rm eff} = 8500\pm$500 K) and evolutionary tracks of mass-transferring binaries, we found current masses of the gainer M$_{2} = 8.8\pm$0.5 M$_{\odot}$ and donor M$_{1} = 0.75\pm0.25$ M$_{\odot}$. We also modeled the object's IR-excess and derived a dust mass of $\sim 5\,\times10^{-5}$ M$_{\odot}$ in the optically-thin dusty disk. The orbital parameters and properties of the H$\alpha$ line profile suggest that the circumstellar gaseous disk is predominantly circumbinary. The relatively low mass of the gainer led us to a suggestion that 3\,Pup should be excluded from the B[e] supergiant group and moved to the FS\,CMa group.
Overall these results further support our original suggestion that FS\,CMa objects are binary systems, where an earlier mass-transfer caused formation of the circumstellar envelope.
\end{abstract}

\keywords{Stars: emission-line, Be; (Stars:) binaries: spectroscopic; Stars: individual: 3\,Pup}

\section{Introduction} \label{intro}
The object 3\,Pup (HD\,62623, HR\,2996) is the brightest ($V \sim$ 4.0 mag) among those exhibiting the B[e] phenomenon. The latter is defined as the presence of emission lines, including forbidden, in the spectra of B-type stars as well as an IR excess due to dust radiation. The emission-line spectrum of the star was first detected by \citet{1934PASP...46..156M}. The first identification of spectral lines in a wavelength range from 3300 \AA\, to H$\alpha$ was published by \citet{1950AnAp...13..114S}. A more detailed history of spectral studies of 3\,Pup along with a line list found at a high resolution ($R = 60,000-70,000$) in a range from 3682 to 8863 \AA\, can be found in \citet{2010AstBu..65..150C}. An atlas of the spectrum of 3\,Pup in the spectral range from 3920 \AA\ to 6920 \AA\ at $R = 60,000$ is presented by \citet{2015AstBu..70...99K}. Parts of the spectrum at longer wavelengths from 7280 \AA\ to 7340 \AA\ and 8480 \AA\ to 8680 \AA\ at $R =15,000 - 18,000$ are presented by \citet{2016MNRAS.456.1424A}.

Regular radial velocity (RV) variations of the absorption lines in the spectrum of 3\,Pup were first detected by \citet{1946PASP...58..248J}, who reported an orbital period of 137.767 days and a RV semi-amplitude of 3.60$\pm$0.45 km\,s$^{-1}$. These photographic data with addition of several higher-resolution CCD spectra were reanalyzed by \citet{1995A&A...293..363P}. Using a Fourier analysis technique, these authors found two equally probable orbital periods of 138.5 and 161.1 days. They also estimated the object's distance modulus to be 10.8--12.4 mag corresponding to distances from 1.4 to 3.0 kpc. This result led \citet{1995A&A...293..363P} to a very large luminosity of the star ranging from $(1.1 \pm 0.5) \times 10^5$ L$_{\odot}$ and a conclusion that 3\,Pup was a massive star. Such a large distance and a high luminosity were not supported by the results of \citet{2010AstBu..65..150C}, who derived a distance of 650 pc from the spectral classification of the star (A2.7 {\sc i}b) and comparison of its spectrum with those of other early A-type supergiants.

The IR excess of 3\,Pup was analyzed by \citet{1994MNRAS.266..203R,1995A&A...293..363P,2004ApJ...602..978S,2010A&A...512A..73M}. The former authors attributed the excess to chromospheric emission, while the others clearly showed that it is due to radiation of the circumstellar dust. A strong emission band, indicative of optically-thin radiation of silicate dusty particles was detected in the low-resolution IRAS spectrum. \citet{2004ApJ...602..978S} calculated a model SED of the star and the gas-and-dust disk-like envelope using a stellar temperature of 10,000 K, a typical accelerated stellar wind of a single supergiant with a latitudinal mass flux distribution, and spherical silicate dusty grains. The best model found by these authors contained dusty grains with a diameter of 1 micron and was consistent with near-IR interferometric data.

More recent studies \citep[e.g.,][]{2004ApJ...605..436M,2010A&A...512A..73M,2011A&A...526A.107M} reported IR interferometry of the object, resolved the dusty disk around it, and concluded that the material accumulation in its equatorial plane is most likely due to the presence of a low-mass component. Also, spectroscopic studies by \citet{2015ApJ...800L..20K} and \citet{2016MNRAS.456.1424A} in the visual and near-IR regions suggested a Keplerian motion of the material in the circumstellar disk around the A-supergiant.

Nevertheless, there is still much uncertainty in the stellar and circumstellar parameters of the object. In particular, there is no generally accepted agreement on the luminosity of the A-supergiant, mass function of the binary, and even the orbital period.
In this paper we report the results of a long-term high-resolution spectroscopic monitoring of 3\,Pup and a new attempt to model the dusty portion of the system IR excess with the goal to further constrain the system parameters.

\section{Observations}\label{observations}

Spectroscopic observations of 3\,Pup were obtained by A.M. and S.D. in 2012--2020 at the Three College Observatory in North Carolina (131 spectra). Additionally 17 spectra were obtained in 2004--2018 at four additional observatories. Table\,\ref{t1} summarizes these observations.

\begin{table*}[htb]
\caption[] {Observatories and instruments used for spectroscopic observations}
\begin{center}
\begin{tabular}{lllcl}
\hline\noalign{\smallskip}
Observatory & Telescope & Instrument & Resolution                       & Location \\
ID                 &                  &                   & $\lambda/\Delta\,\lambda$&               \\
\noalign{\smallskip}\hline\noalign{\smallskip}
TCO                  & 0.81\,m at the Three College   & eShel (Shelyak                                                  &  12,000   & North Carolina, USA\\
                          & Observatory                            & Instruments)\footnote{http://www.shelyak.com} &                &                                  \\
\noalign{\smallskip}\hline\noalign{\smallskip}
SAO                  & 6\,m at the Special Astrophysical         & NES                                                      & 60,000    & Nizhniy Arkhyz, Russia\\
                          & Observatory of the Russian Academy & \citep{2017ARep...61..820P}                 &                &                                      \\
                          & of Sciences                                          &                                                               &                &                                      \\
\noalign{\smallskip}\hline\noalign{\smallskip}
OAN                  & 2.1\,m at the Observatorio Astron\'omico & REOSC                                             & 18,000    & Baja California, Mexico\\
SPM                  & Nacional San Pedro Martir                   &                                                               &                &                                      \\
\noalign{\smallskip}\hline\noalign{\smallskip}
McD                   & 2.7\,m Harlan J. Smith                         &Tull coud\'e TS2                                     & 60,000  & Mt. Locke, Texas, USA \\
                          & at the McDonald Observatory             &   \citep{1995PASP..107..251T}                &                &                                      \\
\noalign{\smallskip}\hline\noalign{\smallskip}
CFHT                & 3.6\,m Canada-France-Hawaii            & ESPaDOnS                                              & 65,000 & Mauna Kea, Hawaii, USA\\
                          & Telescope                                           & \citep{2003SPIE.4843..425M}                  &                &                                      \\
\noalign{\smallskip}\hline\noalign{\smallskip}
\end{tabular}
\end{center}
\label{t1}
\end{table*}

The data obtained at OAN SPM, McDonald (McD), and TCO were reduced in a standard way using the $echelle$ task of IRAF.
Observations obtained at CFHT were reduced with the Upena and Libre-ESpRIT software packages \citep{1997MNRAS.291..658D}.
Typical uncertainties in the wavelength calibration are $<$ 0.5\,km\,s$^{-1}$ for the CFHT, McDonald, and TCO data and $\sim$1 km\,s$^{-1}$ for OAN SPM.
RV standard stars were observed every night at TCO and SPM to control the wavelength calibration. Unexpected deviations from the regular RV variations described in Sect.\,\ref{orbit} were detected only several times and might have been due to either the pulsational activity of the A--supergiant or to effects of flexure (the spectrograph at OAN SPM is mounted at the Cassegrain focus).
Log of the observations along with the measurements of the H$\alpha$ line properties and cross-correlated RV is presented in Table\,\ref{t2}.

\begin{table*}[htb]
\caption[] {Log of spectroscopic observations of 3\,Pup}
\begin{center}
\begin{tabular}{lcrlcrccccc}
\hline\noalign{\smallskip}
Date          &HJD$-$2450000& Obs.  &Range, \AA            & Phase            &  RV        &$\sigma$(RV)& I$_{\rm V}$ & I$_{\rm R}$ & I$_{\rm d}$ & $V/R$ \\
\noalign{\smallskip}\hline\noalign{\smallskip}
12/25/04	         &	3365.042	&    CFHT	&	3600$-$10500	&	0.287	&$-$5.5	&	0.3	&	1.92	&	2.84	&	0.84	&	0.68	\\
12/12/06$^b$	&	4082.943	&	SPM	&	3850$-$6865	&	0.511	&$-$		&	$-$	&	1.95	&	2.31	&	1.24	&	0.84	\\
11/15/07	         &	4420.016	&	SPM	&	3600$-$6775	&	0.964	&$-$1.1	&	0.4	&	1.63	&	2.12	&	1.36	&	0.77	\\
11/20/07$^b$	&	4424.982	&	SPM	&	3600$-$6775	&	0.000	&$-$		&	$-$	&	1.65	&	2.30	&	1.48	&	0.72	\\
10/10/08	        &	4750.017	&	SPM	&	3690$-$6800	&	0.365	&$-$4.4	&	0.5	&	1.75	&	2.98	&	1.15	&	0.59	\\
11/04/08$^a$     &	4774.627	&	SAO	&	4462$-$5926	&	0.545	&$-$0.5	&	0.1	&	$-$	&	$-$	&	$-$	&	$-$	\\
12/11/08$^c$  &	4811.881	&	McD	&	3600$-$10140	&	0.816	&$-$		&	$-$	&	1.60	&	2.88	&	1.14	&	0.56	\\
\noalign{\smallskip}\hline\noalign{\smallskip}
\end{tabular}
\end{center}
\begin{list}{}
\item Log of spectroscopic observations of 3\,Pup. Full Table is shown in the electronic version of the paper.
Column information: (1) -- Calendar date (MM/DD/YY), (2) -- Julian Date (JD$-$2450000), (3) - Observatory ID (see Table\,\ref{t1}), (4) -- spectral range observed,
(5) -- orbital phase according to the RV solution (see text), (6) -- radial velocity in km\,s$^{-1}$ derived by cross-correlation in the range 4460--4632 \AA\, (see Fig.\,\ref{f1}), (7) uncertainty in the radial velocity determination in km\,s$^{-1}$,
(8--11) -- parameters of the H$\alpha$ line profiles: blue peak intensity in continuum units (I$_{\rm V}$), red peak intensity (I$_{\rm R}$), intensity of the central depression (I$_{\rm d}$), and the peak intensity ratio ($V/R$).
\item Comments on the spectra with no RV measurements: $^a$ -- a large portion or the entire cross-correlation region was not observed;
$^b$ -- a systematic error in the wavelength calibration;
$^c$ -- region damaged by reflection on the CCD chip.
\end{list}
\label{t2}
\end{table*}

\begin{figure}[t]
\begin{center}
\includegraphics[width=9.0cm, bb = 20 20 730 530, clip=]{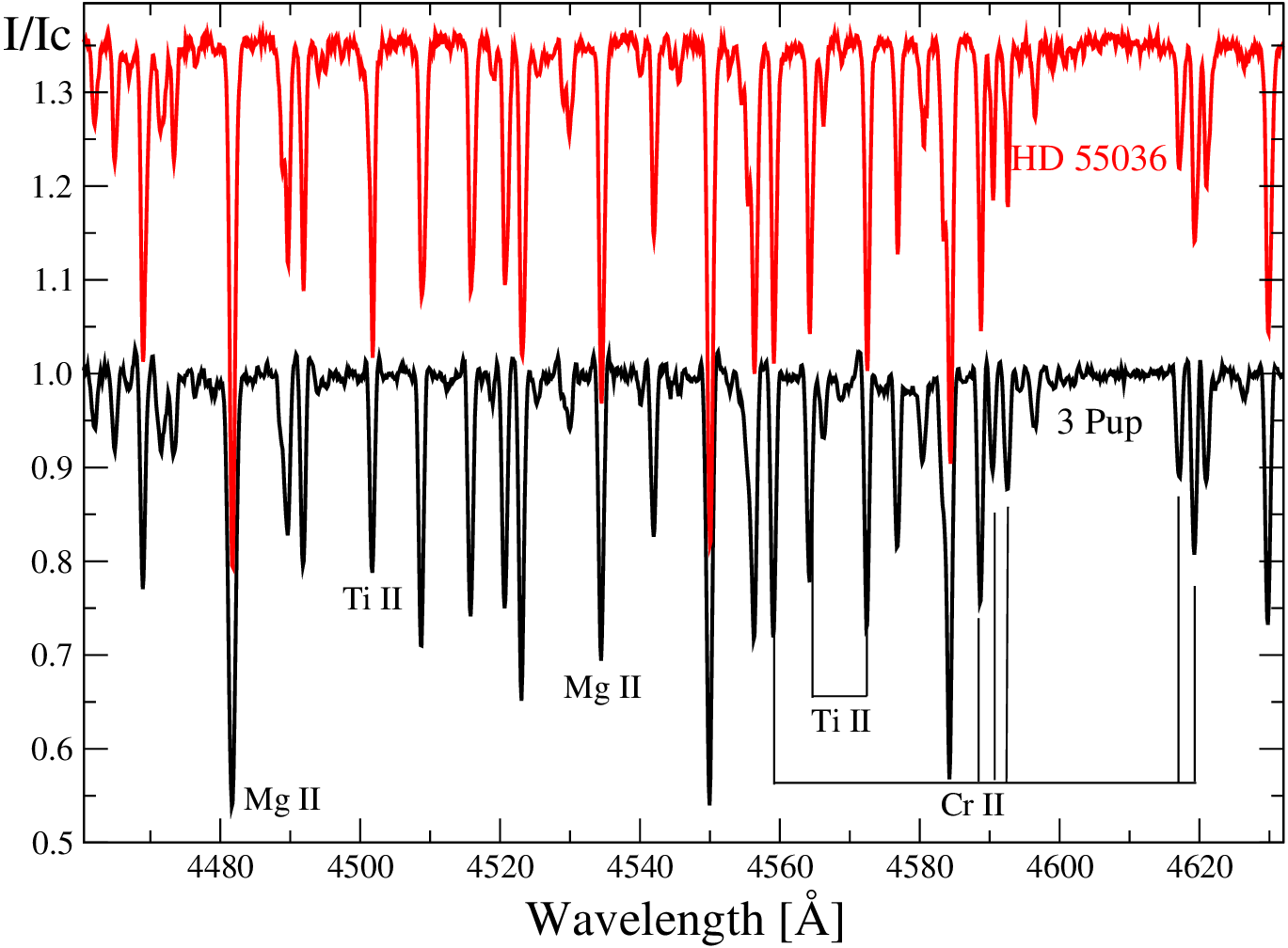}
\end{center}
\caption{Template for cross-correlation of the absorption spectrum of 3\,Pup  taken at TCO (lower spectrum). Part of the spectrum of HD\,55036, an A3 {\sc i}b supergiant, taken at the 2\,m OAN SPM telescope and convolved with the TCO spectral resolution is shown for comparison. Intensity is normalized to the local continuum, the wavelength scale is shifted to match the line positions and given in Angstroems. Identification of some lines is shown. Most other lines belong to Fe {\sc ii}.
}
 \label{f1}
\end{figure}

\section{Data Analysis}\label{analysis}

\subsection {Binary Orbit}\label{orbit}

The optical spectrum of 3\,Pup and line identification were described in detail in \citet{2010AstBu..65..150C}. Our goal in this study was to investigate long-term line position and profile variations using a much larger data set than those reported previously. To reach this goal, we have measured the emission peaks intensity ratios in the H$\alpha$ profile and positions of over 30 absorption lines of Fe {\sc ii}, Mg {\sc ii}, Ti {\sc ii}, and Cr {\sc ii} in a blue region from 4460 \AA\, to 4632 \AA\, as well as in a red region around the Si {\sc ii} lines at 6347 \AA\, and 6371 \AA. Spectra with both regions of absorption lines were normalized to a local continuum and cross-correlated against a template spectrum, which was chosen to be a TCO spectrum of 3\,Pup obtained on February 17, 2017 (signal to noise ratio in continuum over 200, see Fig.\,\ref{f1}), using the $rvsao$ package in IRAF. The template spectrum was obtained near an orbital phase, at which the observed RV is close to the systemic velocity. Also, using an individual spectrum as a template is better than an averaged one to avoid a spectral line broadening because of additional variations due to other processes, such as pulsations (see a brief discussion below).

The spectrum of HD\,55036 shown for comparison in Fig.\,\ref{f1} was selected from a sample of A2--A3 {\sc i}b supergiants that have been observed at TCO and OAN SPM. We show it just to demonstrate that the spectrum of 3\,Pup looks like that of a typical supergiant in the regions with no line emission.

\begin{figure}[t]
\begin{center}
\begin{tabular}{l}
\includegraphics[width=8.0cm, bb = 20 20 730 530, clip=]{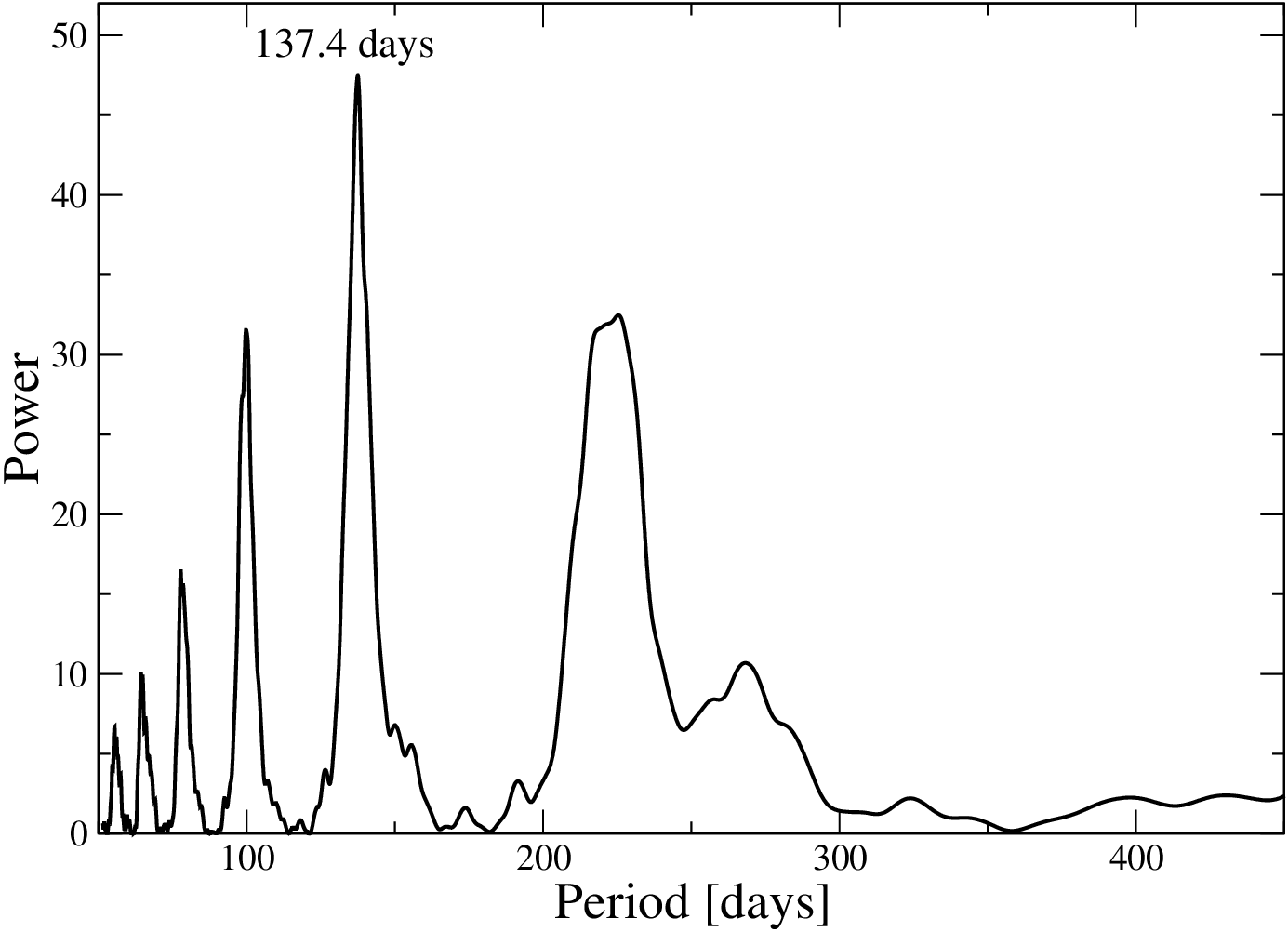}\\
\includegraphics[width=8.0cm, bb = 20 20 730 530, clip=]{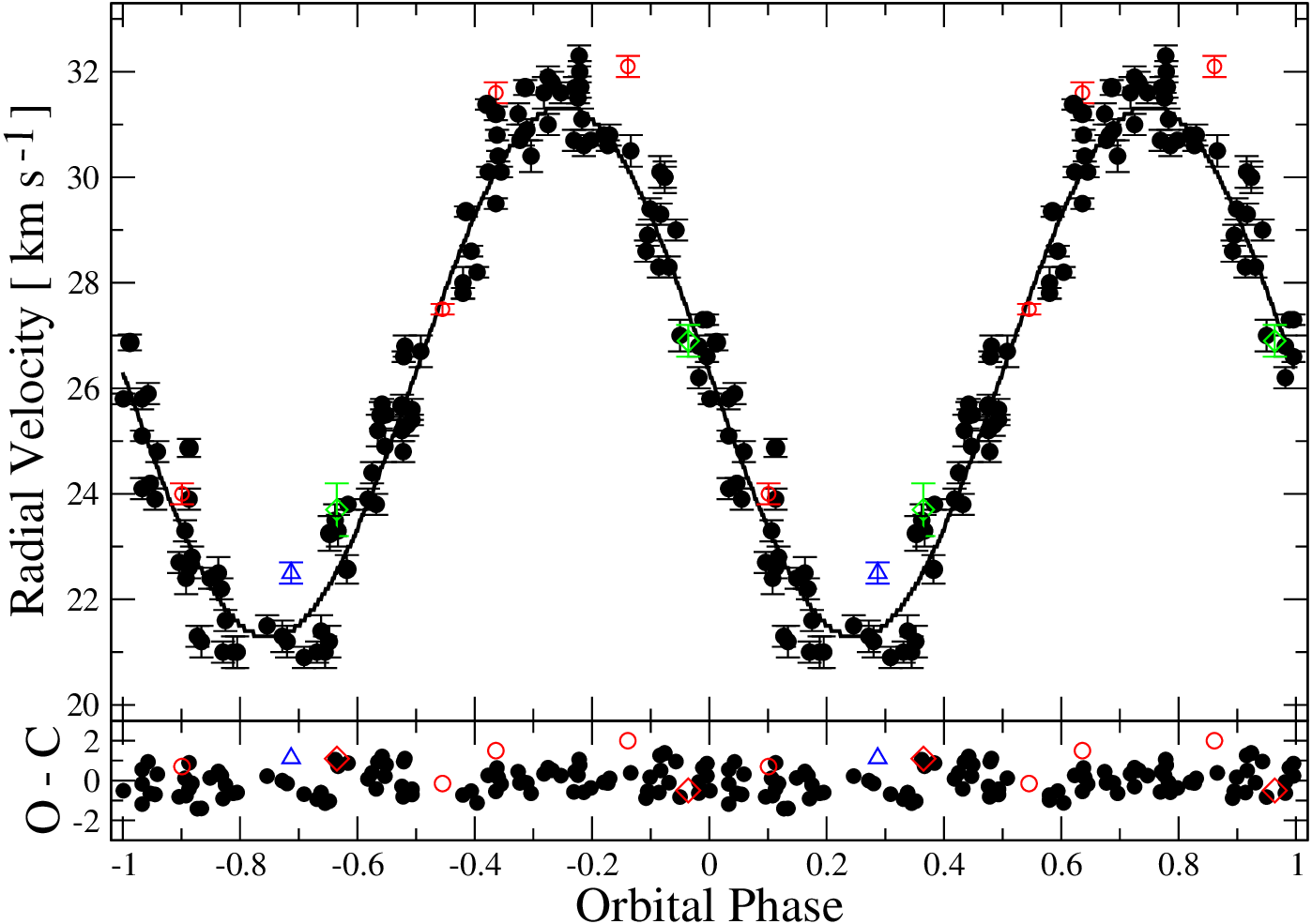}\\
\end{tabular}
\end{center}
\caption{
{\bf Upper panel:} Fourier power spectrum for the RV variations.
{\bf Lower panel:} RV phase curve based on cross-correlation of the 4460--4632 \AA\, spectral region. The RV scale is heliocentric and shown in km\,s$^{-1}$. Filled circles represent TCO data, open circles show SAO data, diamonds show SPM data, and the triangle shows CFHT data. Bottom section of the plot shows deviations of the observed RV from calculated one.
}
 \label{f2}
\end{figure}

The main result found from the cross-correlation is a very well-defined pattern of the RV variations (see Fig.\,\ref{f2}). It turned out to be sinusoidal within the measurement uncertainties, indicating a circular orbit of the A-supergiant around the center of mass of the system.
The best fit to the blue region data has the following parameters: the orbital period P$_{\rm orb} = 137.4\pm0.1$ days, RV semi-amplitude $K = 5.0\pm0.8$ km\,s$^{-1}$, and the epoch of the superior conjunction HJD$_{0}$ = HJD245,3325.7$\pm$3.0. The systemic velocity $\gamma = +26.4\pm2.0$ km\,s$^{-1}$ was determined by adding the average RV from cross-correlation and the average RV of all the absorption lines in the blue region of the template spectrum.

Although the RV measurement accuracy is of the order of 0.3 km\,s$^{-1}$, the uncertainty of the derived RV semi-amplitude due to the orbital motion is nearly three times higher. It is most likely due to short-term variations of the line positions mentioned by \citet{2010AstBu..65..150C}. Such variations may be caused by pulsations typical of early-type supergiants \citep[e.g.,][]{2013MNRAS.433.1246S}.

The RV curve folded with the orbital period is shown in the lower panel of Fig.\,\ref{f2}.
The cross-correlation of the spectra in the red region showed virtually the same results as for the blue region.
The orbital period and the RV semi-amplitude result in a mass function $f(m) = (1.81^{+0.97}_{-0.76}) \times 10^{-3}$ M$_{\odot}$ which is discussed in Sect.\,\ref{discussion}.

\subsection{Other Spectral Variations}\label{variations}

Previous papers that reported high-resolution spectra of 3\,Pup contained only limited discussion of the line profile variations because of a small number of spectra obtained. For example, \citet{2010AstBu..65..150C} mentioned profile variations of Fe {\sc ii} and Na {\sc i} lines comparing only two spectra. Our collection allows tracing variations of spectral lines over a period of nearly two decades. Overall, the absorption-line spectrum of 3\,Pup in the visual region is typical of {\sc i}b supergiants, and absorption lines show very weak variations of their profiles, taking into account uncertainties due to continuum normalization and wavelength calibration.

\begin{figure}[t]
\includegraphics[width=8.0cm, bb = 20 20 730 530, clip=]{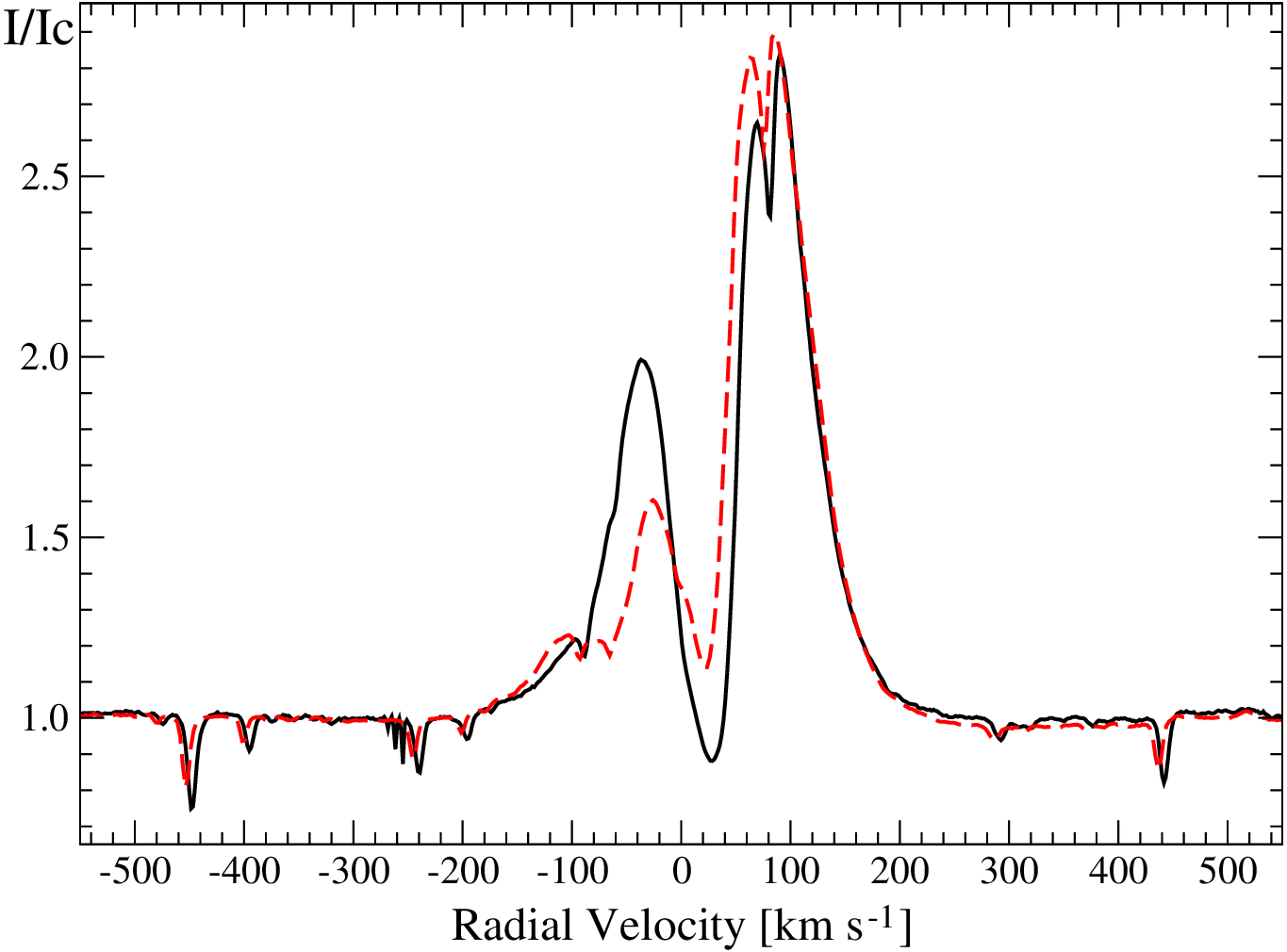}
\caption{H$\alpha$ line profiles in the SAO spectrum taken on 2013 October 14 (solid line) and the McDonald spectrum taken on 2008 December 11 (dashed line). Both spectra have the same $R$ = 60,000. Telluric lines were not removed from the data. Intensity is normalized to the local continuum, the RV is shown in km\,s$^{-1}$.}
 \label{f3}
\end{figure}

We analyzed the H$\alpha$ line double-peaked profile variations and found several features of its long-term behavior. The first one concerns the strength of the central depression, which varied between 1.08 and 1.48 of the continuum intensity even in our highest-resolution spectra in 2006--2012. Since October 2013 it has never exceeded 1.03 of the continuum intensity. As seen in Fig.\,\ref{f3}, the blue-shifted peak was lower during the former period compared to the latter one. Also, the blue-shifted peak was accompanied by a weaker and even bluer-shifted peak at a RV of $\sim -100$ km\,s$^{-1}$.

\begin{figure*}[htb]
\vspace*{0.5cm}
\begin{center}
\begin{tabular}{ll}
\vspace*{0.1cm}\resizebox*{0.45\hsize}{7.0cm}{\includegraphics{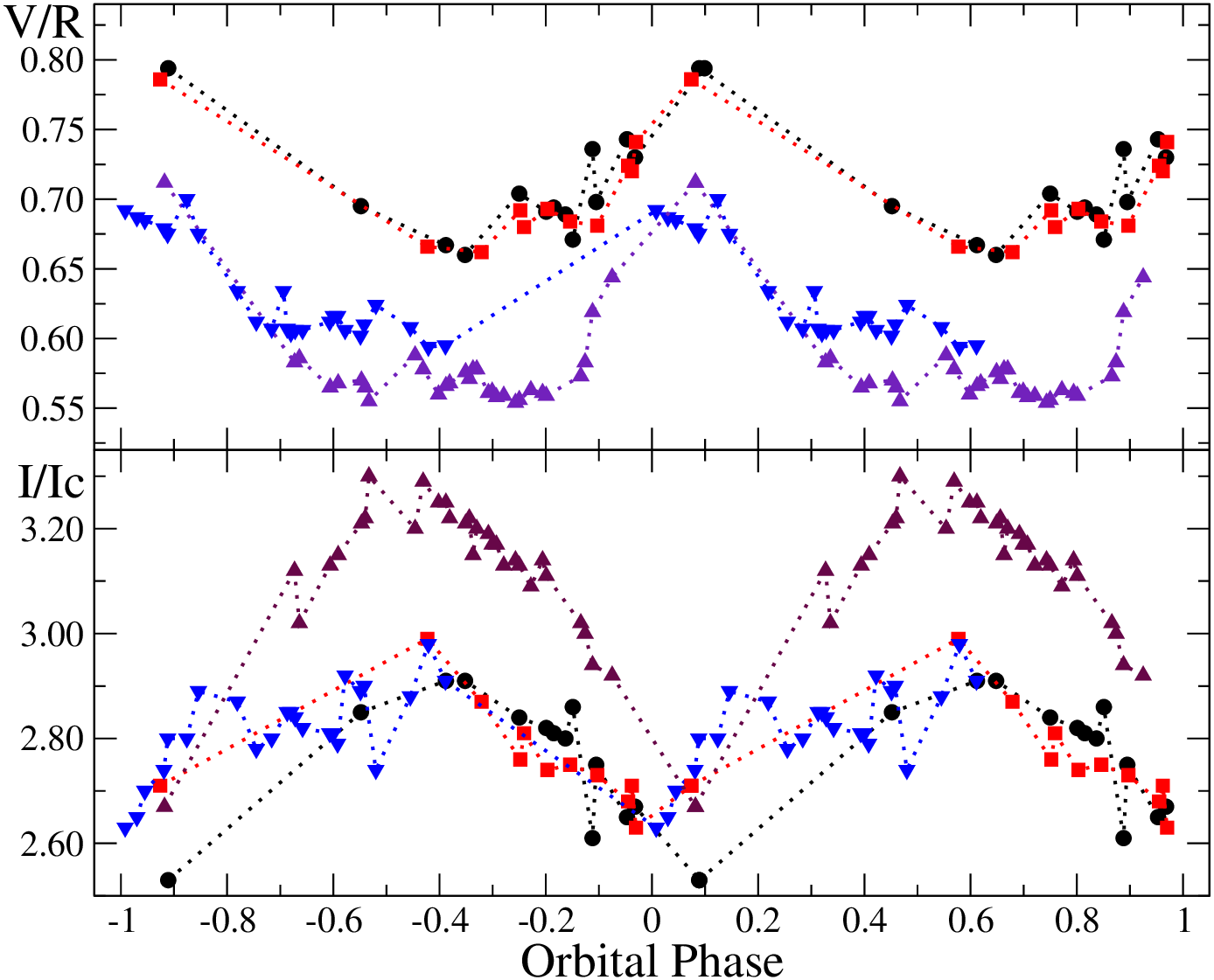}}&
\hspace*{0.6cm}\resizebox*{0.45\hsize}{7.0cm}{\includegraphics{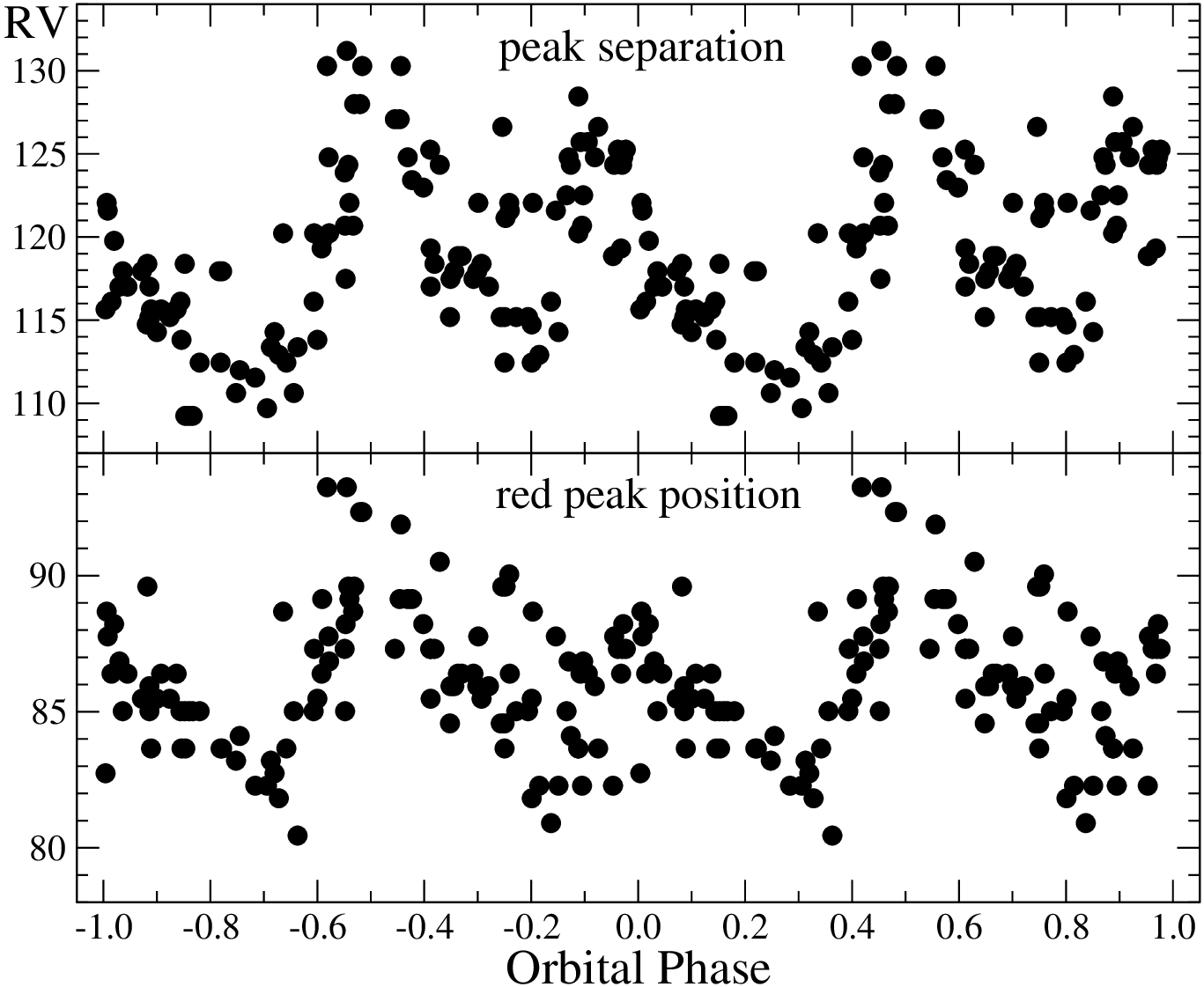}}\\
\end{tabular}
\end{center}
\caption{Variations of the H$\alpha$ line profile parameters folded with the orbital phase. {\bf Upper left panel:} peak intensity ratio. {\bf Lower left panel:} red emission peak. Intensity scale is given in the continuum units. Dotted lines connect the measurements taken during the same orbital cycle: 2014 -- circles, 2015 -- squares, 2017 -- upward triangles, 2018 -- downward triangles. Only the cycles with a representative phase coverage are shown.\\
{\bf Upper right panel:} The H$\alpha$ line emission-peak separation. {\bf Lower right panel:} The H$\alpha$ line red emission peak position. RVs are shown in km\,s$^{-1}$.}
\label{f4}
\end{figure*}

Other features show correlation with the orbital phase (see Fig.\,\ref{f4}). The peak intensity ratio $V/R$ shows a tendency for a lower $V/R$ near inferior conjunction (orbital phase = 0.5), when the A-supergiant is located in front of the secondary component. At the same time, the red emission peak is noticeably stronger around this phase. The values of $V/R$ and the red peak intensity vary from cycle to cycle, but overall the dependence keeps its shape. Also, the emission peak separation and the red peak position (measured by fitting an area around the peak with a Gaussian) exhibit a well-defined maximum near the same phase.

\subsection{Spectral Energy Distribution}\label{sed}

The IR excess in the SED of 3\,Pup has been discussed in several papers. \citet{1995A&A...293..363P} mentioned that the dusty envelope extends from $R_{in} = 35 R_{\star}$ to $R_{out} = 5,000 R_{\star}$, where $R_{\star}$ is the radius of the A-supergiant, with a radial density profile of $\rho \propto {\rm r}^{-1.3}$ but have not given the modeling details.
The best model of \citet{2004ApJ...602..978S} has similar parameters ($R_{in} = 20 R_{\star}, R_{out} = 5,000 R_{\star}$, and $\rho \propto {\rm r}^{-2.0}$), uses a grain size of 1 $\mu$m, and IR fluxes at wavelengths up to 100 $\mu$m. The latter authors and \citet{2010A&A...512A..73M} focused on explaining the mid-IR interferometry results.

\begin{figure}
\begin{center}
\includegraphics[width=8.0cm, bb = 50 20 710 550, clip=]{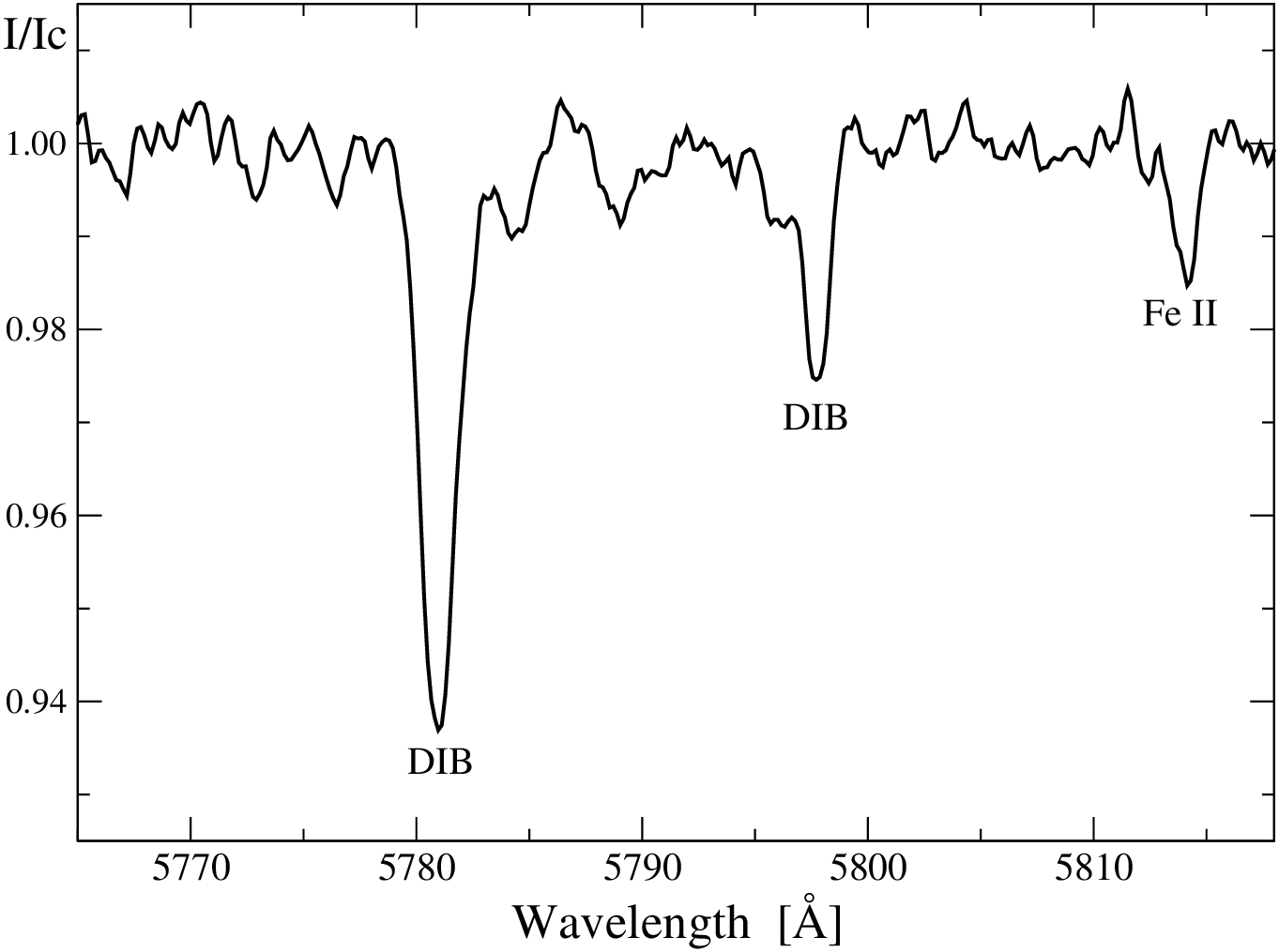} 
\caption{Part of the TCO spectrum of 3\,Pup showing the diffuse interstellar bands (DIBs) that were used to estimate the amount of interstellar reddening. Intensities and wavelengths are in the same units as in Fig.\,\ref{f1}.}
 \label{f5}  
\end{center}
\end{figure}

We tried reproducing the SED shape in a wider wavelength range.
The SED of 3\,Pup was constructed from various sources, which included UV, optical photometry, near-IR, and IRAS photometry \citep[presented in][]{1995A&A...293..363P}, fluxes from more recent IR surveys (WISE, AKARI), and a set of submillimetric fluxes from \citet{2001ApJ...550L..71J}. The star has been found to show little variability in the visual spectral range \citep[0.03 mag, e.g.,][]{1993A&AS..102...79S,1997IBVS.4541....1A}. Therefore, combining data taken at different times to construct the SED is justified. The SED was corrected for the interstellar extinction using $E(B-V) = 0.15$ mag (see Fig.\,\ref{f5}) and the average Galactic reddening law from \citet{1979ARA&A..17...73S}.

It was assumed that the IR excess radiation at wavelengths $\lambda \ga 1 \mu$m originates from a circumbinary dusty disk.
The flux density in this case is given by

\begin{equation}\label{form7}
 f_{\nu,disk}= d^{-2} \int_{R_{in}}^{R_{out}}B_{\nu}(T_r) (1 - exp(-\tau_{\nu,r})) 2\pi r dr,  
 \end{equation}
where $R_{in}$ and $R_{out}$ are the disk inner and outer radii, respectively, $d$ is the distance to the star, and $B_{\nu} (T)$ is the Plank function. The computations were done for a distance of 650~pc. Optical depth of the disk material $\tau_{\nu,r}$ is the product of a wavelength-dependent disk opacity, $\kappa_{\nu}$, and the radial surface density distribution, $\Sigma_r$. To calculate the disk opacity, we used the Mie theory and considered spherical grains composed of astronomical silicates with a density of 2.5~g\,cm$^{-3}$, sizes of $a = 0.01-100~\mu$m \citep[see][for a detailed description of the grain emissivity determination]{2013MNRAS.431.1573B}, and a size distribution $\propto a^{-3.5}$ typical of interstellar grains \citep{1977ApJ...217..425M}. We assumed that the disk has a gas to dust mass ratio of 100 \citep{1978ApJ...224..132B}. Contributions from both gas and dust are taken into account in calculating the opacity of the circumstellar material.

\par The temperature $T_r$, vertical height $H_r$ and surface density $\Sigma_r$ distributions were parametrized as power laws of the disk inner radius:
\begin{equation}\label{form10}
  T_r=T_{in}\left(\frac{r}{R_{in}}\right)^{-\tilde{q}},
    \end{equation}
    
 \begin{equation}\label{Hr}
  H_r=H_{in}\left(\frac{r}{R_{in}}\right)^{\frac{3-\tilde{q}}{2}},
    \end{equation}   
    
\begin{equation}\label{form9}
  \Sigma_r=\Sigma_{in}\left(\frac{r}{R_{in}}\right)^{-p},
    \end{equation}
where $T_{in}$ is the disk temperature at $R_{in}$, that we calculate using the radiative equilibrium equation.
The index $\tilde{q}$ and the disk vertical height at the inner edge $H_{in}$ are taken to be free parameters. The index $p$ describes the surface density variation with distance from the star.
$\Sigma_{in}$ is a surface density at $R_{in}$. The total mass of the disk is related to the disk size and surface density as follows:
\begin{equation}\label{md}
M_d = \int_{R_{in}}^{R_{out}}2 \pi r~\Sigma_r~{\rm d}r.
\end{equation}
Conversely, one can express $\Sigma_{in}$ as a function of the disk mass:
\begin{equation}\label{Sigma_r}
\Sigma_{in} = \frac {M_d~R_{in}^{-p}~(2-p)}{2 \pi~(R_{out}^{2-p} - R_{in}^{2-p})}.
\end{equation}
Geometry of a disk tilted with respect to the line of sight is accounted for by the approach described in \citet{2015AASP....5...33Z}. It assumes that the disk edges at the inner and outer radii have a cutoff flat geometry and emit as blackbodies with constant temperatures derived from the radiative equilibrium equation and Eq.~\ref{form10} at the inner and outer edges, respectively.\\

\par Using the modeling approach described above and tested on the SED of the young star IRAS\,22150+6109 \citep{2018MNRAS.477..977Z}, a grid of SEDs for the disk with different parameters was calculated. The ranges and increments of the modeling parameters are shown in Table\,\ref{t3}.

\par
\begin{figure}
\begin{center}
\includegraphics[width=8.0cm, bb = 25 42 710 550, clip=]{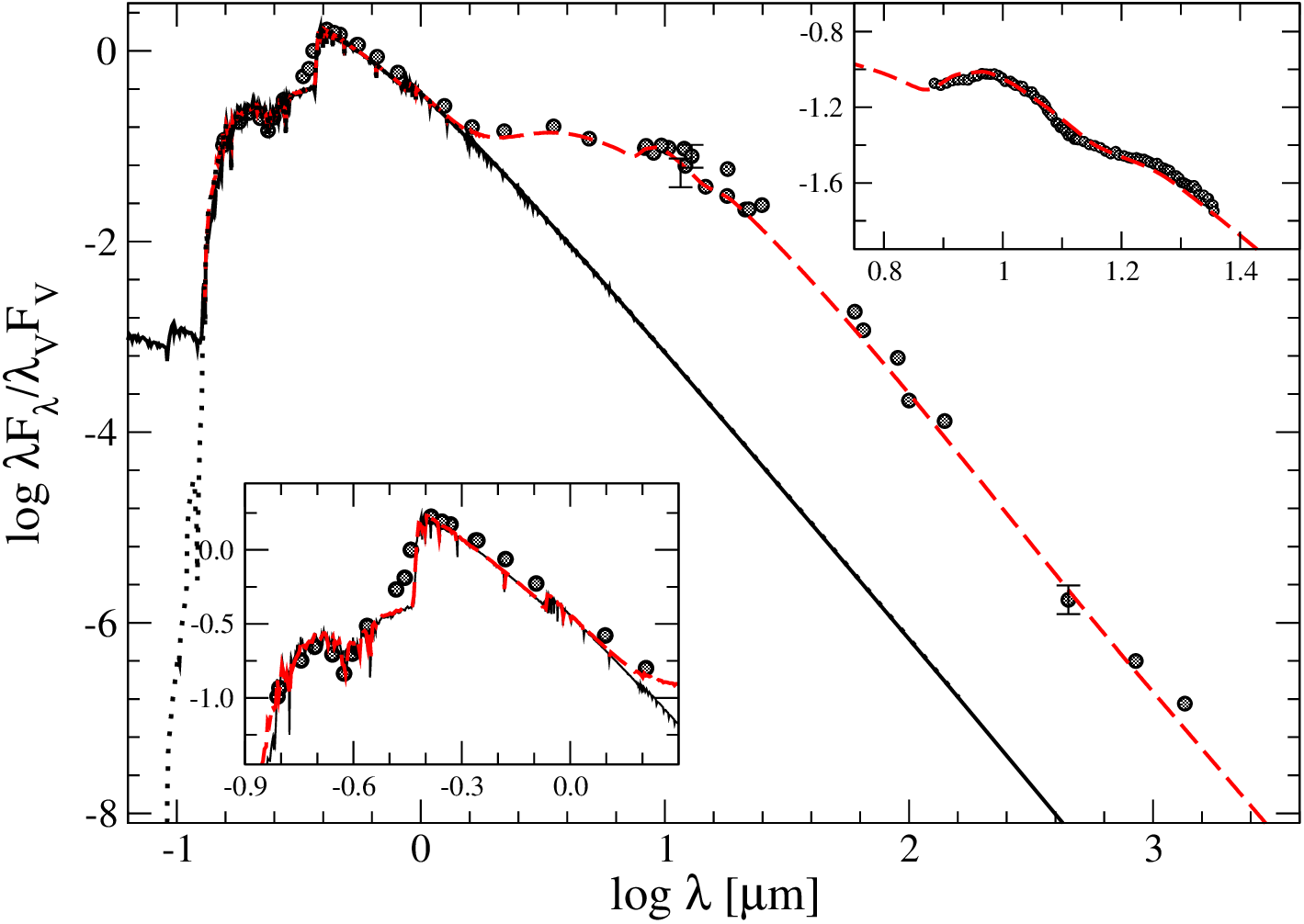} 
\caption{The best-fit disk model (dashed line) is shown in comparison with the interstellar reddening corrected observed SED of 3\,Pup composed from the mentioned photometric data (filled circles) and IRAS low resolution spectrum (open circles in the upper inset). The lower inset shows the agreement of the observed and model UV fluxes. The uncertainties of the observational data points do not exceed the symbol size except for those at $\lambda =11.6, 12.8$, and 450 $\mu$m, whose errors are $\sim 30 \%$. The photospheric radiation of the binary is shown with the solid line. It is represented by the scaled sum of the emergent fluxes from the mass gainer with T$_{\rm eff}$ = 8500 K and $\log$ g = 2.0 and the mass donor with T$_{\rm eff}$ = 50,000 K and $\log$ g = 5.0 from a model grid by \citet{2003IAUS..210P.A20C} for the solar element abundances. The component contributions are set by parameters from Table\,\ref{t4}. The flux from the mass donor becomes dominant at wavelengths shorter than $\sim$ 1000 \AA. The dotted line shows a model SED of the mass gainer alone.}
 \label{f6}  
\end{center}
\end{figure}

 The best fit for the SED was found by minimizing:
 
\begin{equation}\label{formX2}
  \chi^2=\sum^{n}_{i=1}\Big({\frac{F_{obs,i}-F_{mod,i}}{F_{obs,i}}}\Big)^2,
    \end{equation}
where $F_{obs,i}$ and $F_{mod,i}$ are the observed and modeled fluxes (at the corresponding wavelength) respectively. We normalize all the differences to the observed fluxes to account for their large range ($\sim 10^6$, see Fig.\,\ref{f6}). We compared the modeled and observed SED only at wavelengths $>$~1~$\mu$m, where the disk dominates the object's radiation. At all tilt angles used in the fitting process, there was no attenuation of the A-supergiant by the disk. This is justified by a very good agreement of the dereddened observed UV fluxes and the intrinsic model SED for the chosen T$_{\rm eff}$ of the star as well as by a much poorer agreement of the observed and model fluxes in the near IR region at a very high tilt angle of 80$\degr$.

We fixed $R_{in}$ at 4~AU, because this is consistent with the results of the interferometric observations \citep{2010A&A...512A..73M,2011A&A...526A.107M}. This value is in agreement with the minimum possible radius, which is the dust sublimation radius (3.88~AU), assuming that the dust sublimation temperature is 1500~K. 
The best-fit ($\chi^2$ = 1.6, Fig.\,\ref{f6}) model parameters are listed in the second column of Table\,\ref{t3}. Three of them were constrained reasonably well: $H_{in} = 0.08^{+0.02}_{-0.05} R_{in}$, $\tilde{q}$ = 0.75$^{+0.05}_{-0.15}$ and $p$ = 1.5$^{+0.3}_{-0.5}$. The error bars correspond to 1~$\sigma$ uncertainties calculated from the $\Delta\chi^2$ confidence statistics.

Since the disk is optically-thin, which is supported by the presence of the silicate emission features at 10 and 18 $\mu$m (see the upper inset in Fig.\,\ref{f6}), the theoretical SED is not very sensitive to the tilt angle variation in a range of 40$\degr$--60$\degr$ and still result in a reasonably good fit to the observed one.
The fits get worse outside of this angle range. Therefore, our results on the tilt angle are in good agreement with those from the previous modeling: $60\degr \pm 10\degr$ by \citet{2010A&A...512A..73M} and $38\degr$ \citet{2011A&A...526A.107M}.

Similarly, the model does not constrain the disk outer radius, $R_{out}$, which determines the flux at the longest wavelengths. The minimum value of $R_{out}$ that gives a reasonably good agreement with the observed sub-mm flux is 50 AU. Increasing it to 120 AU only weakly contributes to the long-wavelength flux given the temperature decrease with distance from the star and a low optical depth of the dust. This uncertainty affects the total disk mass, which becomes an order of magnitude larger that the best-fit value presented in Table~\ref{t3} at $R_{out} \sim 120$ AU. Nevertheless, even with these problems the total mass of the circumstellar material (dust and gas) remains below 0.01 M$_{\odot}$ and represents a very small fraction of the total mass of the entire system.

\begin{table}[!h]
\caption[]{Parameters of the dusty disk}
\begin{center}
\begin{tabular}{llll}
\hline\noalign{\smallskip}
Parameter      & Best-fit                            & Range                                           & Increment\\
\noalign{\smallskip}\hline\noalign{\smallskip}
$R_{\rm out}$ & 60~AU                                  & 20 $\div$ 850 AU                          & 10 AU       \\
$\tilde{q}$       &   0.75                                    & $0.00 \div 0.85$                            & 0.05          \\
$p$                 &   1.5                                      & $0.0 \div 2.0$                                & 0.3            \\
$M_{\rm d}$   &  $5\,10^{-5}~M_{2}$              & $10^{-5} \div 10^{-1} M_{2}$        & 10$\%$     \\
$H_{\rm in}$   &   $0.08\,R_{\rm in}$              & $10^{-3}\div 10^{-1}R_{\rm in}$    & 10$\%$     \\
\noalign{\smallskip}\hline\noalign{\smallskip}
\end{tabular}
\end{center}
\begin{list}{}
\item The mass of the mass gainer, M$_{2}$, is listed in Table\,\ref{t4}. The uncertainties of the best-fit parameters are discussed in the text.
\end{list}
\label{t3}
\end{table}
 
\section{Discussion}\label{discussion}

Our orbital solution clearly suggests a stable and unique orbital period of 137.4 days. This is the same as suggested in the very first study of the RV variations by \citet{1946PASP...58..248J} and does not support later results by \citet{1995A&A...293..363P}. The RV semi-amplitude is small and, along with the orbital period, results in a small mass function (see Sect.\,\ref{orbit}). The mass function depends on three parameters: masses of both components and the tilt angle of the system rotational axis with respect to the line of sight: $f(m) ={{{\rm M}_{1} \times \sin^{3} i} \over {(1 + q^2)}}$,
where M$_1$ is the mass of the currently lower-mass component, $q$ = M$_2$/M$_1$ is the component's mass ratio, M$_2$ is the mass of the A-supergiant, and $i$ is the tilt angle.

In order to estimate the mass of the A-supergiant, its fundamental parameters need to be determined. The spectral type has varied from B8 {\sc i} to A3 {\sc ii} in different papers summarized in {\it SIMBAD}, while the effective temperature was reported to be from 8250 K \citep{2016MNRAS.456.1424A} to 10000 K \citep{2004ApJ...602..978S}. Comparison with optical spectra of a number of A--type supergiants done by \citet{2010AstBu..65..150C} and in this study (see Fig.\,\ref{f1}) favors T$_{\rm eff}$ = 8500 K with an uncertainty on the order of 500 K. It is also consistent with the UV and visual SED corrected for the interstellar reddening $E(B-V) = 0.15\pm0.02$ mag, which was derived from spectral features, such as the $\lambda$2175 \AA\ band \citep{1994MNRAS.266..203R} and diffuse interstellar bands (see Fig.\,\ref{f5}), whose EW were converted into E($B-V$) using a calibration by \citet{1993ApJ...407..142H}.

The interstellar extinction calculated from the derived reddening, A$_{V} = 3.1 \times 0.15 = 0.47$ mag, the absolute visual magnitude of M$_{V} = -5.5\pm0.3$ mag from \citet{2010AstBu..65..150C}, and the average visual magnitude ($V = 3.96\pm0.03$ mag) imply a distance of 630$\pm$85 pc. The latter within the measurement uncertainty coincides with the one (633$^{+160}_{-107}$ pc) calculated from the GAIA parallax \citep[1.58$\pm$0.32 mas,][]{2018A&A...616A...1G}. A very similar distance (650 pc) was used in all the papers published after 2004 quoted here.

\begin{figure}[t]
\setlength{\unitlength}{1mm}
\resizebox{9.cm}{!}{
\begin{picture}(70,54)(0,0)
\put (-2,2) {\includegraphics[width=6.95cm,  clip=]{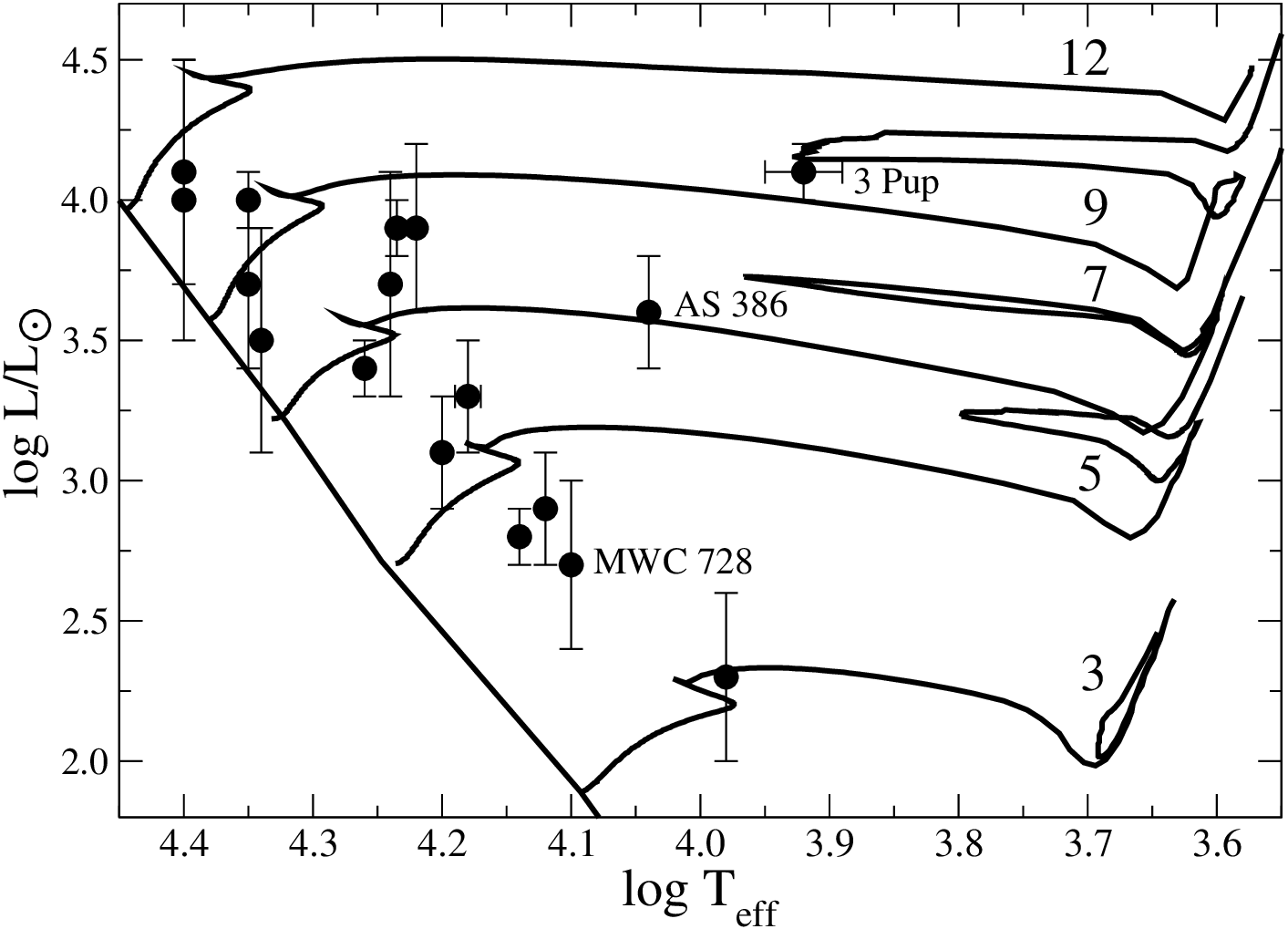}}
\end{picture}}
\caption{{Hertzsprung-Russell diagram with positions of FS\,CMa objects with known fundamental parameters \citep[c.f.][]{2007ApJ...667..497M}. Evolutionary tracks for single rotating stars are taken from \citet{2012A&A...537A.146E} and shown by solid lines along with the main-sequence location. Numbers by the tracks indicate initial masses in solar units. Recently identified binary systems are marked with their IDs.
\label{f7}}}
\end{figure}

The mass of the A-supergiant can be estimated from the evolutionary tracks for single stars, as no signs of the current mass transfer are seen in both photometric (no large brightness variations) and spectroscopic data (weak emission-line spectrum with small and gradual line profile variations).
The adopted values of M$_{V}$ and T$_{\rm eff}$ imply a luminosity of $\log$ L/L$_{\odot} = 4.1\pm0.1$.
Comparison of these parameters with the evolutionary tracks for rotating single stars with initial atmospheric solar abundances \citep{2012A&A...537A.146E}  suggests two possible values of the A-supergiant mass: 9.5$\pm$0.5 M$_{\odot}$ for the pre-red-supergiant phase and 8.8$\pm$0.5 M$_{\odot}$ for the post-red-supergiant phase (see Fig.\,\ref{f7}).

\begin{figure}[t]
\begin{center}
\includegraphics[width=8.0cm, bb = 30 30 720 530, clip=]{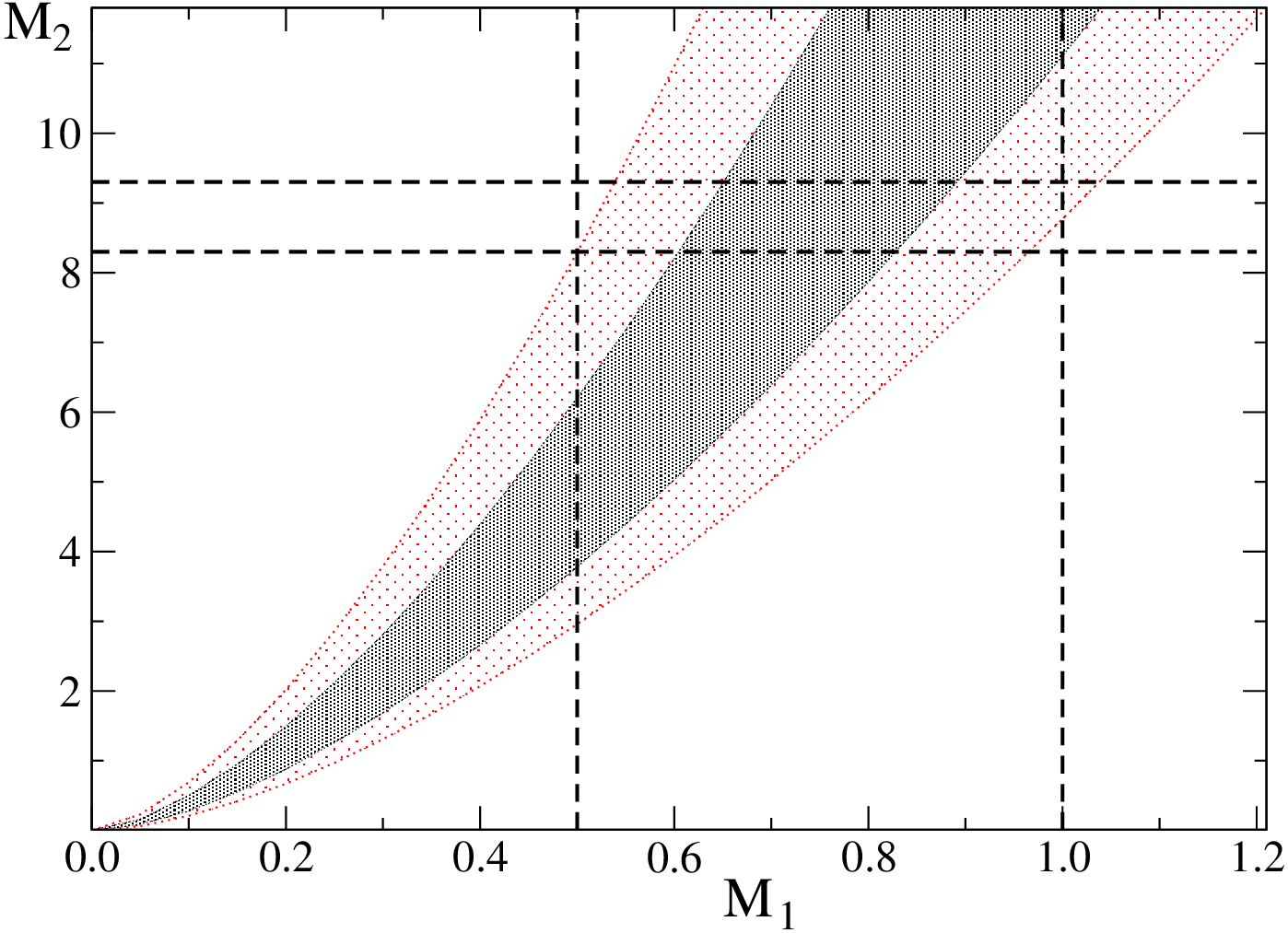}
\end{center}
\caption{Mass function of the 3\,Pup binary system. M$_1$ and M$_2$ are masses of the donor and gainer in solar units, respectively.
The darker-shaded area represents the relationship between the masses for the orbital tilt angles of 40$^{\degr}$ and 60$^{\degr}$ and the
best-fit mass function ($f(m) = 1.81\,\times10^{-3}$ M$_{\odot}$). The lighter-shaded area shows how the relationship changes for $f(m) \pm 1\,\sigma$
within the same interval of tilt angles.
Vertical dashed lines show suggested limits for the mass of the donor based on the probable range of tilt angles. Horizontal dashed lines show the limits on the mass of the gainer from Table\,\ref{t4}.
}
\label{f8}
\end{figure}

These mass estimates along with the orbital solution can be used to derive the mass of the secondary component. 
The dusty disk tilt angle derived from our SED modeling is consistent with the interferometric results \citep{2010A&A...512A..73M,2011A&A...526A.107M}.
Assuming that the binary orbit is tilted with respect to the line of sight at the same angle as the circumbinary disk ($i = 40\degr-60\degr$ from both interferometric and IR-excess modeling data), the secondary component should have a mass of M$_{1} \sim 0.8$ M$_{\odot}$ (see Fig.\,\ref{f8}).

There can be two evolutionary scenarios that lead to the current masses of the stars in this system. One scenario assumes the components' evolution as single stars with no mass-exchange, while the other one involves mass-transfer between the components.

In the former scenario, the lower-mass component would be a solar-like star, which is much fainter than the A-supergiant, still located at main-sequence and not contributing to the observed spectrum. However it seems to be less likely, because the A-supergiant is not massive enough to develop a strong stellar wind, which would supply sufficient amount of material for the observed gaseous and dusty disk. Spectroscopic observations of many supergiants typically show either P\,Cyg type profiles of Balmer lines (in the most luminous ones with the luminosity type {\sc i}a) or no line emission at all (in less luminous ones). Even the most luminous ones do not show the presence of dust in their circumstellar environments \citep[e.g.,][]{1999A&A...346..819V}. 

This reasoning applies to the earlier possible evolutionary phase of the A-supergiant (pre-red-supergiant). The possibility that it is currently a post-red-supergiant is very unlikely.  When it reaches the lower-temperature end of the evolutionary track, its radius becomes so large that the system will go through a common envelope phase given the orbital period. After this phase the element abundances of the brighter component will be significantly altered \citep[e.g., the hydrogen abundance becomes much lower, c.f.][]{1998A&ARv...9...63V}, while the observed absorption spectrum of 3\,Pup is typical for a pre-red-supergiant phase star (see Fig.\,\ref{f1}). Therefore, we favor the scenario of the system evolution with mass-transfer between the components.

The derived components' masses and the presence of the circumstellar material are qualitatively consistent with the evolutionary models of binary systems that undergo mass transfer calculated by \citet{2008A&A...487.1129V}.
A strong mass transfer during the Roche lobe overfilling time by a more massive and more evolved donor resulted in transferring most of its mass to the gainer as well as losing a fraction of the transferred mass to the circumstellar medium and beyond. Some models of \citet{2008A&A...487.1129V} show that the orbital period can reach several months after the end of the mass transfer phase, and that even the mass donor may get down to such a low mass as 0.6--1.0 M$_{\odot}$. 

\begin{figure}[t]
\setlength{\unitlength}{1mm}
\resizebox{8.7cm}{!}{
\begin{picture}(70,54)(0,0)
\put (-2,2) {\includegraphics[width=6.95cm,  clip=]{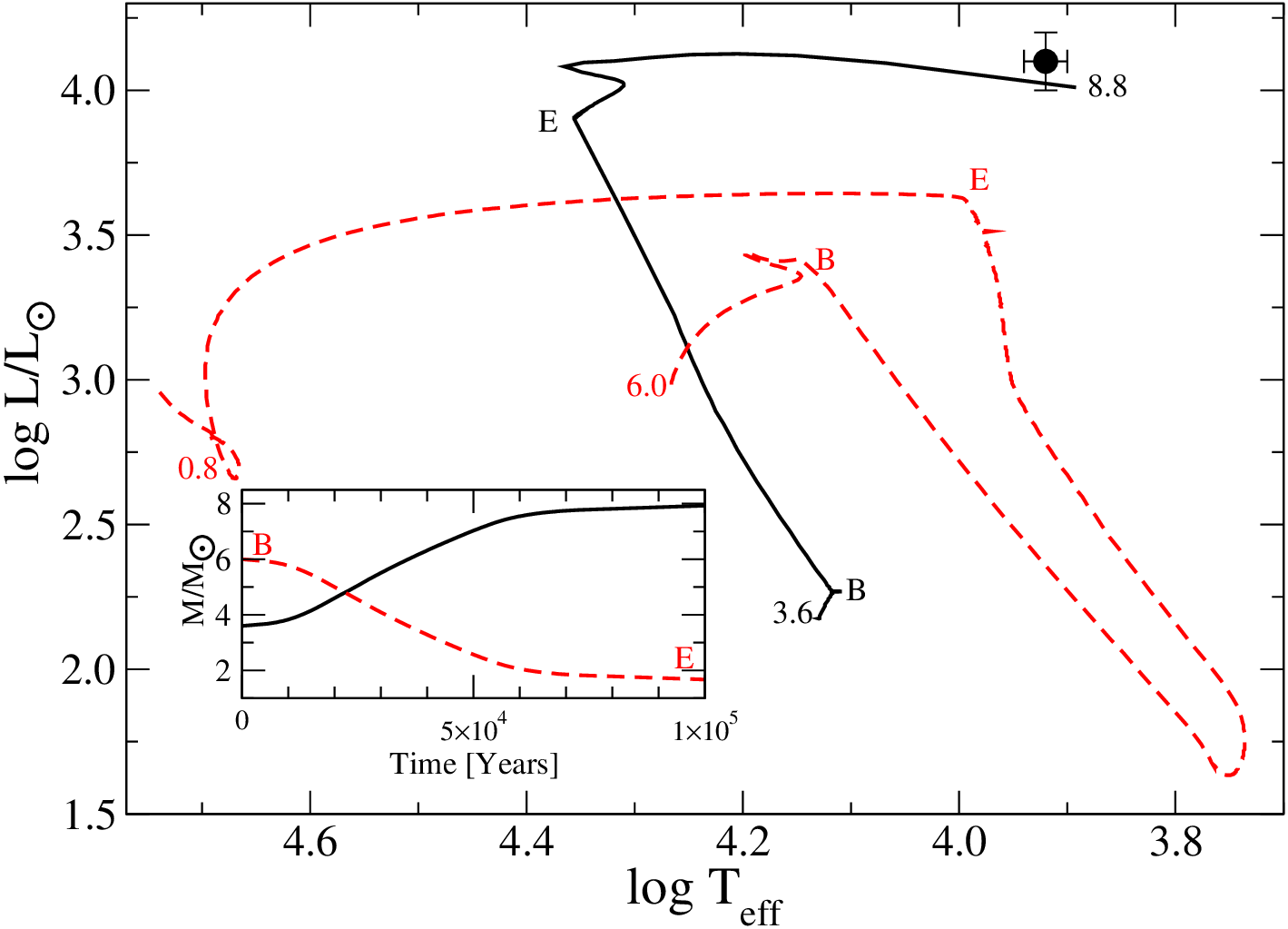}}
\end{picture}}
\caption{{Theoretical evolutionary tracks of a 6.0+3.6 M$_{\odot}$ binary system with an initial orbital period of 5 days that we calculated using the code described in \citet{2008A&A...487.1129V}. The dashed line shows the track for the mass donor, while the solid line show that of the mass gainer. The numbers near the tracks indicate initial and final masses of the components in solar units, while the letters indicate the beginning and ending of the Roche overflow phase. 
The black circle shows the current position of the mass gainer of the 3\,Pup binary system. The inset shows the mass transfer process in relative time since the beginning of Roche lobe overflow phase.
\label{f9}}}
\end{figure}

Since there was no close match to the system parameters in the published model grid, we extended it using the same calculation approach and found one that has the final period (134 days) and luminosity of the more massive component close to those derived above. The adopted evolutionary track has initial masses of 6.0+3.6 M$_{\odot}$ and an initial orbital period of 5.0 days using the same approach. It assumes a conservative mass transfer, since the mass of the circumstellar material that we found in our SED modeling (Sect.\,\ref{sed}) is very small (see Fig.\,\ref{f9}). The Roche lobe overflow phase (partially shown in the inset of Fig.\,\ref{f9}) lasts for 0.4 Myrs. The system is currently 9 Myrs after this phase, $\sim$10 \% of its total evolutionary time. These results along with a lack of studies of more distant B- and A-type supergiants may explain a rare detection of such objects.

The evolutionary track for the mass gainer gets close to the current position of the A-supergiant with a mass of 8.8 M$_{\odot}$. After the end of the mass transfer phase, which lasts for $\sim 10^5$ years, the mass gainer evolves as a single star with nearly the final mass (with a small mass loss through the stellar wind). The mass donor has a final mass of 0.8 M$_{\odot}$, a T$_{\rm eff} \sim 50,000$ K, and a bolometric luminosity an order of magnitude smaller than that of the mass gainer. With these parameters, the hot He-rich dwarf would noticeably contribute to the binary's SED only at wavelengths below 1000 \AA\, (see Fig.\,\ref{f6}). The spectrum of 3\,Pup has not been observed in this far-UV region yet.

The difference in the single and binary evolutionary mass of the gainer is not large ($\sim$10 \%) and can be attributed to differences in the evolutionary codes of \citet{2012A&A...537A.146E} and \citet{2008A&A...487.1129V} as well as to a different internal structure of the star that is set by the evolutionary path. Since the object's binarity was revealed and confirmed, we adopt the binary evolutionary mass of M$_{2} = 8.8\pm$0.5 M$_{\odot}$ for the gainer. The mass uncertainty is based on that of the luminosity and should be similar to that estimated from the single star evolutionary tracks. The gainer's mass leads to a slightly altered estimate for the donor's mass (M$_{1} = 0.75\pm$0.25 M$_{\odot}$, see Fig.\,\ref{f8}).

\begin{table}[!h]
\caption[]{Parameters of the system components}
\begin{center}
\begin{tabular}{lll}
\hline\noalign{\smallskip}
Parameter      & Gainer                           & Donor\\
\noalign{\smallskip}\hline\noalign{\smallskip}
T$_{\rm eff}$     & 8500$\pm$500 K & $\sim$50,000 K \\
R/R$_{\odot}$.  & 54$\pm$7            & $\sim$0.3 \\
$\log$ g            & 1.9$\pm$0.1         & $\sim$5.0 \\
M/M$_{\odot}$ & 8.8$\pm$0.5         & 0.75$\pm$0.25\\
\noalign{\smallskip}\hline\noalign{\smallskip}
\end{tabular}
\end{center}
\begin{list}{}
\item The parameters are based on our analysis of the spectrum of the A-supergiant and the evolutionary model of the binary (Fig.\,\ref{f9}). T$_{\rm eff}$, R, $\log$ g, and M are the current effective temperature, radius, surface gravity, and mass of each of the system components.
\end{list}
\label{t4}
\end{table}

Spectroscopically, the mass gainer in the 3\,Pup binary system is a supergiant and has always been considered belonging to the B$[$e$]$ supergiants subgroup. 
All other objects of the latter subgroup are typically much more massive with an average luminosity of an order of magnitude larger than that of the most luminous FS\,CMa objects \citep[e.g.,][]{2007ApJ...667..497M}. The position of this component in the Hertzsprung-Russell diagram (see Fig.\,\ref{f7}) implies that it is currently more consistent with the system classification as a FS\,CMa object. This subgroup of objects with the B[e] phenomenon was defined by \citet{2007ApJ...667..497M}, and their properties were interpreted as due to consequences of the mentioned above non-conservative binary evolution. Most properties of 3\,Pup (except for the strength of the emission-line spectrum) match those of the FS\,CMa group. 
Thus, we suggest a re-classification of the 3\,Pup binary system into a FS\,CMa object.

The orbital solution and the components' masses result in an orbital semi-major axis of $a = 1.11\pm0.03$ AU or $4.4^{+0.7}_{-0.5}$ R$_{2}$, where R$_{2}$ is the radius of the A-supergiant. According to \citet{1983ApJ...268..368E}, the Roche lobe of the mass gainer has a radius of 0.65\,$a$ for the current components' mass ratio. Therefore, the mass gainer is well confined within its Roche lobe, and its gaseous disk is limited by the Roche lobe size to $\sim$2.9 R$_{2}$.

The derived mass and radius of the mass gainer imply a critical rotation velocity $v_{\rm crit} = 436.8\,\sqrt{{\rm ({M_{2}/M}}_{\odot})/({\rm {R_{2}/R}}_{\odot}}) = 178^{+18}_{-16}$ km\,s$^{-1}$. Our measurements with the Fourier transform method show that the projected rotational velocity of the A-supergiant is $v_{\rm rot} = 35\pm5$ km\,s$^{-1}$. This result indicates a slower rotation compared to the previously reported data \citep[e.g., $50\pm5$ km\,s$^{-1}$,][]{1995A&A...293..363P}.
Thus, one should not expect an enhanced mass loss from it. Assuming that the Keplerian gaseous disk extends all the way to the star and its density does not significantly drop outward, the separation of the line emission peaks should be $v_{\rm peak} = 2\,v_{\rm crit}\,\sin i \sim 273$ km\,s$^{-1}$ \citep[c.f.,][]{1972ApJ...171..549H}. However as seen in Fig.\,\ref{f4}, the average peak separation in the H$\alpha$ emission line is $\sim$120 km\,s$^{-1}$. The latter value is expected at a distance of $\sim$6\,R$_{2}$, near the L2 point of the system. Along with a stable position of the H$\alpha$ line central depression, this result may be interpreted as an evidence for a circumbinary location of the gaseous disk. This idea was suggested by \citet{1995A&A...293..363P}, but the location of the second star in the system was unknown at that time.

The typically weaker blue-shifted peak in this line can then be explained by the presence of a stellar wind from the A-supergiant. The presence of the
wind is expected for such a large and a relatively massive star, but, as we mentioned above, a purely wind-driven H$\alpha$ emission is usually observed only in more luminous in A-type {\sc i}a supergiants \citep[e.g.,][]{1999A&A...346..819V}. Nevertheless, even a weak wind might provide enough mass flux into the disk to distort the observed H$\alpha$ line profile. Modeling of the spectral line profiles is a complicated task and will be attempted in a follow-up paper.

The H$\alpha$ line profile features described in Sect.\,\ref{variations} can be due to the following mechanisms. The higher level of the central depression along with the presence of the weak bluest peak (see Fig.\,\ref{f3}) may be due to an increase in the mass-loss rate from the A-supergiant. This process eventually results in mixing the new material with the disk and, as a consequence, a lower contribution of the circumstellar gas to the RV region responsible for the central depression. The $V/R$ dependence on the orbital phase may be due to a higher contribution of the stellar wind to the line profile, since a smaller fraction of the part of the wind directed toward the observer is attenuated by the mass donor near the superior conjunction orbital phase. The redshift of the red emission peak along with a greater peak separation (see Fig.\,\ref{f4}) do not contradict the above explanation.

\section{Conclusions}

A long-term monitoring of the brightest object exhibiting the B$[$e$]$ phenomenon, 3\,Pup, resulted in refining the orbit of this binary system and detecting variations of the H$\alpha$ line profile, the strongest emission feature in the object's optical spectrum. From the spectroscopic and GAIA parallaxes, which agree with one another, the derived orbital solution, and consideration of the binary system evolutionary models, we determined the current masses of the system components to be M$_{2} = 8.8\pm0.5$ M$_{\odot}$ and M$_{1} = 0.75\pm0.25$ M$_{\odot}$. With such a mass, the mass gainer better fits into the group of FS\,CMa objects rather than into that of B$[$e$]$ supergiants. The mass donor may be a helium-rich subdwarf of a much lower luminosity, explaining why no contribution from such a high-temperature object is observed in the UV fluxes of 3\,Pup.

An evolutionary model of a binary system with the components' initial masses of 6.0 and 3.6 M$_{\odot}$ and conservative mass transfer was calculated to explain the current fundamental parameters of the A--type component and the observed orbital period. Although it does not imply mass loss from the system, the small amount of the circumstellar material estimated from the IR-excess modeling is not expected to noticeably change the components' evolution.  

We also suggest that the circumstellar gaseous disk is circumbinary. This follows from the separations of the double-peaked emission-line profiles. A stellar wind, which is expected from the A-supergiant, can be responsible for the lower strength of the blue-shifted peak in the H$\alpha$ line profile. The circumstellar dust is located further away from the stars than the circumstellar gas, whose outer parts are most likely neutral due to a low temperature of the A-supergiant and a weak contribution of the ionizing radiation from the mass donor. 

Modeling the object's IR excess allowed us to put new constraints on the amount and structure of its dusty disk. The best disk model agrees with the interferometry data on the tilt angle with respect to the line of sight. It also suggests that a dust mass of $\sim 10^{-5}-10^{-4}$ M$_{\odot}$ is currently present in the circumbinary disk.

Continuation of a spectroscopic monitoring is strongly suggested to increase the phase coverage during one orbital cycle to investigate the line profile variations in more detail and try detecting an onset of the next strengthening of the central depression in the H$\alpha$ line profile. Measurements of the system far-UV fluxes (at $\lambda \le 1000$ \AA) are needed to verify the nature of the mass donor.

\acknowledgements
We are grateful to the anonymous referee for useful suggestions that allowed us to improve the paper, especially the data interpretation.
A.~M. and S.~Zh. acknowledge support from DGAPA/PAPIIT Projects IN\,100617 and IN 102120. V.~G.~K. thanks the Russian Foundation for Basic Research for partial support (project 18--02--00029a). M.~P. acknowledges support from the Russian Foundation for Basic Research project 18--02--00554.
The work was partially carried out within the framework of Project No. BR05236322 ``Studies of physical processes in extragalactic and galactic objects and their subsystems'' financed by the Ministry of Education and Science of the Republic of Kazakhstan.
This paper is partly based on observations obtained with the Canada-France-Hawaii Telescope (CFHT) which is operated by the
National Research Council of Canada, the Institut National des Sciences de l$^{\prime}$Univers of the Centre National de la
Recherche Scientifique de France, and the University of Hawaii as well as on observations obtained at the 2.7\,m Harlan J. Smith
telescope of the McDonald Observatory (Texas, USA), and the 2.1\,m of the Observatorio Astron\'omico Nacional San Pedro Martir (Baja California, M\'{e}xico).
The observations at the Canada-France-Hawaii Telescope were performed with care and respect from the summit of Maunakea which is a significant
cultural and historic site.
This research has made use of the SIMBAD database, operated at CDS, Strasbourg, France.

\bibliography{ms_rev1}
\pagebreak
\appendix
\centerline{Log of spectroscopic observations of 3\,Pup}
{\small
\begin{center}
\begin{tabular}{lcrlcrccccc}
\hline\noalign{\smallskip}
Date          &HJD$-$2450000& Obs.  &Range, \AA            &     Phase       &  RV        &$\sigma$(RV)& I$_{\rm V}$ & I$_{\rm R}$ & I$_{\rm d}$ & $V/R$ \\
\noalign{\smallskip}\hline\noalign{\smallskip}
12/25/04	&	3365.042	&    CFHT	&	3600$-$10500	&	0.287	&$-$5.5	&	0.3	&	1.92	&	2.84	&	0.84	&	0.68	\\
12/12/06$^b$&	4082.943	&	SPM	&	3850$-$6865	&	0.511	&$-$		&	$-$	&	1.95	&	2.31	&	1.24	&	0.84	\\
11/15/07	&	4420.016	&	SPM	&	3600$-$6775	&	0.964	&$-$1.1	&	0.4	&	1.63	&	2.12	&	1.36	&	0.77	\\
11/20/07$^b$&	4424.982	&	SPM	&	3600$-$6775	&	0.000	&$-$		&	$-$	&	1.65	&	2.30	&	1.48	&	0.72	\\
10/10/08	&	4750.017	&	SPM	&	3690$-$6800	&	0.365	&$-$4.4	&	0.5	&	1.75	&	2.98	&	1.15	&	0.59	\\
11/04/08$^a$&	4774.627	&	SAO	&	4462$-$5926	&	0.545	&$-$0.5	&	0.1	&	$-$	&	$-$	&	$-$	&	$-$   \\
12/11/08$^c$&	4811.881	&	McD	&	3600$-$10140	&	0.816	&$-$		&	$-$	&	1.60	&	2.88	&	1.14	&	0.56	\\
12/13/08$^c$&	4813.887	&	McD	&	3600$-$10140	&	0.830	&$-$		&	$-$	&	1.58	&	2.90	&	1.15	&	0.55	\\
12/15/08$^c$&	4815.887	&	McD	&	3600$-$10140	&	0.845	&$-$		&	$-$	&	1.57	&	2.89	&	1.15	&	0.54	\\
02/02/10	&	5230.323	&	SAO	&	3817$-$5279	&	0.861	&	4.2	&	0.2	&	$-$	&	$-$	&	$-$	&	$-$   \\
02/03/10$^a$&	5231.335	&	SAO	&	5160$-$6690	&	0.868	&$-$		&	$-$	&	1.83	&	2.88	&	1.17	&	0.64	\\
01/24/12$^a$&	5951.691	&	TCO	&	4581$-$7243	&	0.110	&$-$		&	$-$	&	2.19	&	2.54	&	1.17	&	0.86	\\
01/27/12$^a$&	5954.685	&	TCO	&	4579$-$7749	&	0.132	&$-$		&	$-$	&	2.12	&	2.50	&	1.16	&	0.85	\\
02/12/12$^a$&	5970.638	&	TCO	&	4540$-$7242	&	0.248	&$-$		&	$-$	&	2.10	&	2.59	&	1.14	&	0.81	\\
03/06/12$^a$&	5993.580	&	TCO	&	4600$-$7485	&	0.415	&$-$		&	$-$	&	2.26	&	2.66	&	1.19	&	0.85	\\
03/10/12$^a$&	5997.571	&	TCO	&	4600$-$7485	&	0.444	&$-$		&	$-$	&	2.18	&	2.59	&	1.08	&	0.84	\\
10/14/13	&	6573.655	&	SAO	&	3916$-$6980	&	0.636	&	3.6	&	0.2	&	1.99	&	2.84	&	0.88	&	0.70	\\
12/12/13	&	6639.812	&	TCO	&	4250$-$7900	&	0.118	&$-$5.2	&	0.2	&	2.01	&	2.53	&	0.91	&	0.79	\\
01/31/14	&	6689.678	&	TCO	&	4250$-$7900	&	0.481	&$-$1.2	&	0.2	&	1.98	&	2.85	&	0.92	&	0.70	\\
02/22/14	&	6711.623	&	TCO	&	4250$-$7900	&	0.640	&	2.4	&	0.1	&	1.94	&	2.91	&	0.86	&	0.67	\\
02/27/14	&	6716.620	&	TCO	&	4250$-$7900	&	0.677	&	2.7	&	0.2	&	1.92	&	2.91	&	0.89	&	0.66	\\
03/13/14	&	6730.572	&	TCO	&	4250$-$7900	&	0.778	&	4.3	&	0.2	&	2.00	&	2.84	&	0.89	&	0.70	\\
03/20/14	&	6737.616	&	TCO	&	4250$-$7900	&	0.830	&	4.1	&	0.3	&	1.95	&	2.82	&	0.90	&	0.69	\\
03/22/14	&	6739.551	&	TCO	&	4250$-$7900	&	0.844	&	4.3	&	0.2	&	1.95	&	2.81	&	0.95	&	0.69	\\
03/25/14	&	6742.578	&	TCO	&	4250$-$7900	&	0.866	&	2.5	&	0.3	&	1.93	&	2.80	&	0.88	&	0.69	\\
03/27/14	&	6744.555	&	TCO	&	4250$-$7900	&	0.880	&	3.8	&	0.2	&	1.92	&	2.86	&	0.87	&	0.67	\\
04/01/14	&	6749.531	&	TCO	&	4250$-$7900	&	0.916	&	2.1	&	0.3	&	1.92	&	2.61	&	0.86	&	0.74	\\
04/02/14	&	6750.540	&	TCO	&	4250$-$7900	&	0.924	&	2.0	&	0.3	&	1.92	&	2.75	&	0.83	&	0.70	\\
04/10/14	&	6758.521	&	TCO	&	4250$-$7900	&	0.982	&$-$1.8	&	0.2	&	1.97	&	2.65	&	0.84	&	0.74	\\
04/12/14	&	6760.520	&	TCO	&	4250$-$7900	&	0.996	&	7.1	&	0.2	&	1.95	&	2.67	&	0.82	&	0.73	\\
04/17/14	&	6765.530	&	TCO	&	4250$-$7900	&	0.033	&$-$2.2	&	0.2	&	1.99	&	2.71	&	0.87	&	0.73	\\
12/04/14$^b$&	6996.918	&	SPM	&	3850$-$7270	&	0.717	&$-$		&     $-$	&	1.91	&	2.75	&	0.94	&	0.70	\\
01/18/15	&	7041.705	&	TCO	&	4250$-$7900	&	0.043	&$-$2.1	&	0.2	&	1.92	&	2.58	&	0.96	&	0.74	\\
02/06/15	&	7060.649	&	TCO	&	4250$-$7900	&	0.180	&$-$7.8	&	0.3	&	1.97	&	2.72	&	0.97	&	0.72	\\
02/07/15	&	7061.656	&	TCO	&	4250$-$7900	&	0.188	&$-$7.0	&	0.3	&	1.94	&	2.74	&	0.97	&	0.71	\\
02/08/15	&	7062.654	&	TCO	&	4250$-$7900	&	0.195	&$-$7.0	&	0.3	&	1.93	&	2.72	&	0.96	&	0.71	\\
03/06/15	&	7088.580	&	TCO	&	4250$-$7900	&	0.384	&$-$4.2	&	0.2	&	1.81	&	2.90	&	1.01	&	0.62	\\
03/07/15	&	7089.568	&	TCO	&	4250$-$7900	&	0.391	&$-$8.5	&	0.2	&	1.80	&	2.87	&	1.02	&	0.63	\\
03/15/15	&	7097.543	&	TCO	&	4250$-$7900	&	0.449	&$-$2.5	&	0.1	&	1.89	&	2.93	&	0.98	&	0.65	\\
10/29/15	&	7324.608	&	SAO	&	3948$-$6982	&	0.101	&$-$4.0	&	0.2	&	2.13	&	2.71	&	0.81	&	0.79	\\
01/05/16	&	7393.733	&	TCO	&	4250$-$7900	&	0.604	&	0.3	&	0.1	&	1.99	&	2.99	&	0.90	&	0.67	\\
01/19/16	&	7407.700	&	TCO	&	4250$-$7900	&	0.706	&	4.4	&	0.2	&	1.90	&	2.87	&	0.87	&	0.66	\\
01/29/16	&	7417.684	&	TCO	&	4250$-$7900	&	0.779	&	4.0	&	0.2	&	1.91	&	2.76	&	0.85	&	0.69	\\
01/30/16	&	7418.669	&	TCO	&	4250$-$7900	&	0.786	&	2.6	&	0.2	&	1.91	&	2.81	&	0.86	&	0.68	\\
02/05/16	&	7424.678	&	TCO	&	4250$-$7900	&	0.830	&	2.8	&	0.2	&	1.90	&	2.74	&	0.85	&	0.69	\\
02/11/16	&	7430.658	&	TCO	&	4250$-$7900	&	0.873	&	3.7	&	0.2	&	1.88	&	2.75	&	0.83	&	0.68	\\
02/18/16	&	7437.621	&	TCO	&	4250$-$7900	&	0.924	&	2.0	&	0.2	&	1.86	&	2.73	&	0.86	&	0.68	\\
02/26/16	&	7445.601	&	TCO	&	4250$-$7900	&	0.982	&$-$1.2	&	0.2	&	1.94	&	2.68	&	0.84	&	0.72	\\
02/27/16	&	7446.607	&	TCO	&	4250$-$7900	&	0.989	&$-$0.7	&	0.1	&	1.95	&	2.71	&	0.84	&	0.72	\\
02/28/16	&	7447.606	&	TCO	&	4250$-$7900	&	0.996	&$-$0.7	&	0.2	&	1.95	&	2.63	&	0.82	&	0.74	\\
03/04/16	&	7452.584	&	TCO	&	4250$-$7900	&	0.033	&$-$3.9	&	0.2	&	1.94	&	2.66	&	0.82	&	0.73	\\
03/06/16	&	7454.576	&	TCO	&	4250$-$7900	&	0.047	&$-$3.8	&	0.2	&	1.98	&	2.66	&	0.85	&	0.74	\\
03/15/16	&	7463.565	&	TCO	&	4250$-$7900	&	0.113	&$-$5.4	&	0.2	&	2.02	&	2.68	&	0.79	&	0.75	\\
03/17/16	&	7465.568	&	TCO	&	4250$-$7900	&	0.127	&$-$6.7	&	0.2	&	2.07	&	2.71	&	0.81	&	0.76	\\
\noalign{\smallskip}\hline\noalign{\smallskip}
\end{tabular}
\end{center}
}

{\small
\begin{center}
\begin{tabular}{lcrlcrccccc}
\hline\noalign{\smallskip}
Date          &HJD$-$2450000& Obs.  &Range, \AA            & Phase            &  RV        &$\sigma$(RV)& I$_{\rm V}$ & I$_{\rm R}$ & I$_{\rm d}$ & $V/R$ \\
\noalign{\smallskip}\hline\noalign{\smallskip}
03/18/16	&	7466.548	&	TCO	&	4250$-$7900	&	0.134	&$-$6.9	&	0.3	&	2.11	&	2.76	&	0.84	&	0.76	\\
03/22/16	&	7470.546	&	TCO	&	4250$-$7900	&	0.163	&$-$5.5	&	0.3	&	2.18	&	2.83	&	0.85	&	0.77	\\
03/28/16	&	7476.531	&	TCO	&	4250$-$7900	&	0.207	&$-$4.5	&	0.3	&	2.14	&	2.78	&	0.85	&	0.77	\\
04/02/16$^b$&	7481.515	&	TCO	&	4250$-$7900	&	0.243	&$-$		&	$-$	&	1.94	&	2.73	&	0.99	&	0.71	\\
12/14/16	&	7737.780	&	TCO	&	4250$-$7900	&	0.108	&$-$5.6	&	0.3	&	1.90	&	2.67	&	0.81	&	0.71	\\
01/17/17$^a$&	7771.422	&	SAO	&	4698$-$7782	&	0.353	&$-$		&	$-$	&	1.82	&	3.12	&	0.73	&	0.58	\\
01/18/17	&	7772.668	&	TCO	&	4250$-$7900	&	0.362	&$-$4.5	&	0.3	&	1.77	&	3.02	&	0.81	&	0.59	\\
01/26/17	&	7780.681	&	TCO	&	4250$-$7900	&	0.420	&$-$2.5	&	0.3	&	1.77	&	3.13	&	0.80	&	0.57	\\
01/28/17	&	7782.654	&	TCO	&	4250$-$7900	&	0.435	&$-$2.8	&	0.3	&	1.79	&	3.15	&	0.81	&	0.57	\\
02/03/17	&	7788.671	&	TCO	&	4250$-$7900	&	0.478	&$-$3.2	&	0.2	&	1.83	&	3.21	&	0.81	&	0.57	\\
02/04/17	&	7789.635	&	TCO	&	4250$-$7900	&	0.485	&$-$2.7	&	0.2	&	1.82	&	3.22	&	0.85	&	0.57	\\
02/05/17	&	7790.640	&	TCO	&	4250$-$7900	&	0.493	&$-$2.4	&	0.2	&	1.83	&	3.30	&	0.83	&	0.56	\\
02/17/17	&	7802.618	&	TCO	&	4250$-$7900	&	0.580	&	0.0	&	0.2	&	1.88	&	3.20	&	0.81	&	0.59	\\
02/19/17	&	7804.620	&	TCO	&	4250$-$7900	&	0.594	&	0.6	&	0.1	&	1.90	&	3.29	&	0.81	&	0.58	\\
02/23/17	&	7808.598	&	TCO	&	4250$-$7900	&	0.623	&	2.1	&	0.1	&	1.82	&	3.25	&	0.78	&	0.56	\\
02/25/17	&	7810.609	&	TCO	&	4250$-$7900	&	0.638	&	2.8	&	0.1	&	1.84	&	3.25	&	0.83	&	0.57	\\
02/26/17	&	7811.598	&	TCO	&	4250$-$7900	&	0.645	&	2.1	&	0.1	&	1.83	&	3.22	&	0.82	&	0.57	\\
03/02/17	&	7815.589	&	TCO	&	4250$-$7900	&	0.674	&	3.2	&	0.2	&	1.85	&	3.21	&	0.85	&	0.58	\\
03/03/17	&	7816.597	&	TCO	&	4250$-$7900	&	0.682	&	2.8	&	0.2	&	1.84	&	3.22	&	0.82	&	0.57	\\
03/04/17	&	7817.587	&	TCO	&	4250$-$7900	&	0.689	&	2.9	&	0.2	&	1.82	&	3.15	&	0.83	&	0.58	\\
03/05/17	&	7818.601	&	TCO	&	4250$-$7900	&	0.696	&	2.4	&	0.3	&	1.85	&	3.20	&	0.85	&	0.58	\\
03/08/17	&	7821.612	&	TCO	&	4250$-$7900	&	0.718	&	3.6	&	0.3	&	1.79	&	3.19	&	0.81	&	0.56	\\
03/09/17	&	7822.587	&	TCO	&	4250$-$7900	&	0.725	&	3.9	&	0.2	&	1.78	&	3.17	&	0.80	&	0.56	\\
03/10/17	&	7823.586	&	TCO	&	4250$-$7900	&	0.732	&	3.8	&	0.2	&	1.77	&	3.17	&	0.83	&	0.56	\\
03/12/17	&	7825.585	&	TCO	&	4250$-$7900	&	0.747	&	3.6	&	0.2	&	1.75	&	3.13	&	0.83	&	0.56	\\
03/15/17	&	7828.567	&	TCO	&	4250$-$7900	&	0.769	&	2.7	&	0.2	&	1.74	&	3.14	&	0.83	&	0.55	\\
03/16/17	&	7829.573	&	TCO	&	4250$-$7900	&	0.776	&	3.5	&	0.1	&	1.74	&	3.13	&	0.82	&	0.56	\\
03/19/17	&	7832.558	&	TCO	&	4250$-$7900	&	0.798	&	2.7	&	0.1	&	1.74	&	3.09	&	0.81	&	0.56	\\
03/22/17	&	7835.561	&	TCO	&	4250$-$7900	&	0.820	&	2.8	&	0.1	&	1.76	&	3.14	&	0.82	&	0.56	\\
03/23/17	&	7836.582	&	TCO	&	4250$-$7900	&	0.827	&	2.6	&	0.2	&	1.74	&	3.11	&	0.79	&	0.56	\\
04/01/17	&	7845.517	&	TCO	&	4250$-$7900	&	0.892	&	0.6	&	0.2	&	1.73	&	3.02	&	0.80	&	0.57	\\
04/02/17	&	7846.509	&	TCO	&	4250$-$7900	&	0.899	&	1.4	&	0.2	&	1.75	&	3.00	&	0.79	&	0.58	\\
04/04/17	&	7848.522	&	TCO	&	4250$-$7900	&	0.950	&$-$1.0	&	0.3	&	1.88	&	2.92	&	0.82	&	0.64	\\
12/04/17$^b$&	8092.939	&	SPM	&	3650$-$7315	&	0.693	&	$-$	&	$-$	&	1.80	&	2.76	&	0.95	&	0.65	\\
12/15/17	&	8103.766	&	TCO	&	4250$-$7900	&	0.771	&	3.7	&	0.2	&	1.81	&	2.74	&	0.93	&	0.66	\\
01/01/18	&	8120.734	&	TCO	&	4250$-$7900	&	0.895	&	0.9	&	0.2	&	1.81	&	2.76	&	0.94	&	0.66	\\
01/04/18	&	8123.743	&	TCO	&	4250$-$7900	&	0.917	&	1.3	&	0.2	&	1.84	&	2.74	&	0.96	&	0.67	\\
01/06/18	&	8125.752	&	TCO	&	4250$-$7900	&	0.931	&	0.3	&	0.2	&	1.83	&	2.72	&	0.97	&	0.67	\\
01/15/18	&	8134.693	&	TCO	&	4250$-$7900	&	0.996	&$-$1.4	&	0.1	&	1.83	&	2.61	&	0.93	&	0.70	\\
01/20/18	&	8139.711	&	TCO	&	4250$-$7900	&	0.033	&$-$2.9	&	0.1	&	1.82	&	2.63	&	0.93	&	0.69	\\
01/23/18	&	8142.712	&	TCO	&	4250$-$7900	&	0.055	&$-$4.1	&	0.2	&	1.82	&	2.65	&	0.91	&	0.69	\\
01/25/18	&	8144.714	&	TCO	&	4250$-$7900	&	0.069	&$-$1.7	&	0.2	&	1.85	&	2.70	&	0.94	&	0.69	\\
01/30/18	&	8149.679	&	TCO	&	4250$-$7900	&	0.106	&$-$4.7	&	0.2	&	1.86	&	2.74	&	0.94	&	0.68	\\
01/31/18	&	8150.664	&	TCO	&	4250$-$7900	&	0.113	&$-$4.1	&	0.2	&	1.89	&	2.80	&	0.97	&	0.68	\\
02/05/18	&	8155.673	&	TCO	&	4250$-$7900	&	0.149	&$-$5.6	&	0.2	&	1.96	&	2.80	&	0.95	&	0.70	\\
02/08/18	&	8158.653	&	TCO	&	4250$-$7900	&	0.171	&$-$7.0	&	0.2	&	1.95	&	2.89	&	0.92	&	0.68	\\
02/18/18	&	8168.634	&	TCO	&	4250$-$7900	&	0.243	&$-$5.7	&	0.3	&	1.82	&	2.87	&	0.93	&	0.63	\\
02/22/18	&	8172.617	&	TCO	&	4250$-$7900	&	0.272	&$-$6.7	&	0.3	&	1.74	&	2.49	&	0.90	&	0.70	\\
02/23/18	&	8173.615	&	TCO	&	4250$-$7900	&	0.280	&$-$6.8	&	0.3	&	1.70	&	2.78	&	0.90	&	0.61	\\
02/27/18	&	8177.595	&	TCO	&	4250$-$7900	&	0.309	&$-$7.1	&	0.3	&	1.70	&	2.80	&	0.93	&	0.61	\\
03/02/18	&	8180.600	&	TCO	&	4250$-$7900	&	0.331	&$-$7.0	&	0.3	&	1.70	&	2.68	&	0.94	&	0.63	\\
03/03/18	&	8181.615	&	TCO	&	4250$-$7900	&	0.338	&$-$6.6	&	0.3	&	1.73	&	2.85	&	0.96	&	0.61	\\
\noalign{\smallskip}\hline\noalign{\smallskip}
\end{tabular}
\end{center}
}

{\small
\begin{center}
\begin{tabular}{lcrlcrccccc}
\hline\noalign{\smallskip}
Date          &HJD$-$2450000& Obs.  &Range, \AA            & Phase            &  RV        &$\sigma$(RV)& I$_{\rm V}$ & I$_{\rm R}$ & I$_{\rm d}$ & $V/R$ \\
\noalign{\smallskip}\hline\noalign{\smallskip}
03/04/18	&	8182.599	&	TCO	&	4250$-$7900	&	0.345	&$-$7.0	&	0.3	&	1.72	&	2.85	&	0.95	&	0.60	\\
03/05/18	&	8183.594	&	TCO	&	4250$-$7900	&	0.352	&$-$6.8	&	0.3	&	1.72	&	2.84	&	0.94	&	0.61	\\
03/07/18	&	8185.590	&	TCO	&	4250$-$7900	&	0.367	&$-$4.7	&	0.3	&	1.71	&	2.82	&	0.97	&	0.61	\\
03/14/18	&	8192.577	&	TCO	&	4250$-$7900	&	0.418	&$-$4.1	&	0.3	&	1.72	&	2.81	&	0.96	&	0.61	\\
03/15/18	&	8193.592	&	TCO	&	4250$-$7900	&	0.425	&$-$3.6	&	0.2	&	1.73	&	2.81	&	0.95	&	0.62	\\
03/16/18	&	8194.580	&	TCO	&	4250$-$7900	&	0.432	&$-$4.2	&	0.2	&	1.72	&	2.79	&	0.98	&	0.62	\\
03/18/18	&	8196.541	&	TCO	&	4250$-$7900	&	0.447	&$-$3.1	&	0.3	&	1.77	&	2.92	&	0.98	&	0.61	\\
03/22/18	&	8200.556	&	TCO	&	4250$-$7900	&	0.476	&$-$2.8	&	0.2	&	1.74	&	2.89	&	0.94	&	0.60	\\
03/23/18	&	8201.542	&	TCO	&	4250$-$7900	&	0.483	&$-$0.7	&	0.2	&	1.77	&	2.90	&	0.98	&	0.61	\\
03/26/18	&	8204.548	&	TCO	&	4250$-$7900	&	0.505	&	2.8	&	0.2	&	1.71	&	2.74	&	0.97	&	0.62	\\
04/04/18	&	8213.514	&	TCO	&	4250$-$7900	&	0.570	&	2.4	&	0.1	&	1.75	&	2.88	&	0.96	&	0.61	\\
04/09/18$^a$&	8218.183	&	SAO	&	4696$-$7780	&	0.604	&	$-$	&	$-$	&	1.77	&	2.98	&	0.88	&	0.59	\\
04/13/18	&	8222.509	&	TCO	&	4250$-$7900	&	0.636	&	1.5	&	0.1	&	1.73	&	2.91	&	0.87	&	0.60	\\
11/20/18	&	8443.848	&	TCO	&	3875$-$7912	&	0.246	&$-$6.5	&	0.2	&	2.00	&	2.54	&	1.01	&	0.79	\\
12/17/18	&	8470.783	&	TCO	&	3860$-$7912	&	0.442	&$-$2.3	&	0.1	&	2.09	&	2.59	&	1.01	&	0.81	\\
12/22/18	&	8475.780	&	TCO	&	3860$-$7912	&	0.479	&$-$1.4	&	0.1	&	2.12	&	2.71	&	1.02	&	0.78	\\
12/24/18	&	8477.744	&	TCO	&	3860$-$7912	&	0.493	&$-$2.6	&	0.2	&	2.07	&	2.65	&	1.01	&	0.78	\\
12/26/18	&	8479.755	&	TCO	&	3860$-$7912	&	0.508	&$-$1.3	&	0.4	&	2.08	&	2.72	&	0.99	&	0.77	\\
01/05/19	&	8489.756	&	TCO	&	3860$-$7912	&	0.580	&$-$0.2	&	0.1	&	1.98	&	2.68	&	0.94	&	0.74	\\
01/15/19	&	8499.723	&	TCO	&	3860$-$7912	&	0.653	&$-$0.2	&	0.1	&	1.96	&	2.69	&	0.95	&	0.73	\\
01/25/19	&	8509.673	&	TCO	&	3860$-$7912	&	0.725	&	3.0	&	0.2	&	1.89	&	2.65	&	0.96	&	0.71	\\
02/02/19	&	8517.664	&	TCO	&	3860$-$7912	&	0.783	&	3.1	&	0.2	&	1.85	&	2.67	&	0.95	&	0.69	\\
02/24/19	&	8539.608	&	TCO	&	3860$-$7912	&	0.943	&	1.0	&	0.2	&	1.94	&	2.62	&	0.95	&	0.74	\\
03/04/19	&	8547.584	&	TCO	&	3860$-$7912	&	0.001	&$-$2.2	&	0.1	&	1.92	&	2.61	&	0.95	&	0.74	\\
03/12/19	&	8555.571	&	TCO	&	3860$-$7912	&	0.059	&$-$3.2	&	0.2	&	1.87	&	2.62	&	0.97	&	0.71	\\
03/17/19	&	8560.552	&	TCO	&	3860$-$7912	&	0.096	&$-$5.4	&	0.3	&	1.84	&	2.65	&	0.96	&	0.69	\\
03/19/19	&	8562.559	&	TCO	&	3860$-$7912	&	0.110	&$-$3.7	&	0.2	&	1.87	&	2.66	&	0.97	&	0.70	\\
03/27/19	&	8570.524	&	TCO	&	3860$-$7912	&	0.168	&$-$5.8	&	0.2	&	1.87	&	2.73	&	0.97	&	0.69	\\
03/28/19	&	8571.528	&	TCO	&	3860$-$7912	&	0.175	&$-$6.4	&	0.2	&	1.85	&	2.54	&	0.94	&	0.73	\\
11/03/19	&	8791.839	&	TCO	&	3860$-$7912	&	0.779	&3.7  	&	0.2	&	1.50	&	2.52	&	1.02	&	0.60	\\
12/05/19	&	8823.836	&	TCO	&	3860$-$7912	&	0.012	&$-$1.1	&	0.2	&	1.99	&	2.49	&	0.97	&	0.80	\\
12/19/19	&	8837.774	&	TCO	&	3860$-$7912	&	0.113	&$-$3.1	&	0.2	&	2.11	&	2.48	&	1.00	&	0.85	\\
01/21/20	&	8870.712	&	TCO	&	3860$-$7912	&	0.353	&$-$4.7	&	0.3	&	1.92	&	2.60	&	1.03	&	0.74	\\
01/25/20	&	8874.690	&	TCO	&	3860$-$7912	&	0.382	&$-$5.4	&	0.3	&	1.91	&	2.64	&	0.99	&	0.72 \\
02/02/20	&	8882.681	&	TCO	&	3860$-$7912	&	0.440	&$-$2.5	&	0.3	&	2.00	&	2.69	&	1.01	&	0.74 \\
02/07/20	&	8887.653	&	TCO	&	3860$-$7912	&	0.476	&$-$2.3	&	0.3	&	2.03	&	2.79	&	1.03	&	0.73 \\
02/22/20	&	8902.636	&	TCO	&	3860$-$7912	&	0.585	&1.4	        &	0.1	&	1.99	&	2.82	&	0.99	&	0.71 \\
02/27/20	&	8907.612	&	TCO	&	3860$-$7912	&	0.621	&3.4	        &	0.1	&	1.99	&	2.80	&	1.00	&	0.71 \\
02/29/20	&	8909.599	&	TCO	&	3860$-$7912	&	0.636	&3.2	        &	0.1	&	2.00	&	2.78	&	1.01	&	0.72 \\
03/07/20	&	8916.577	&	TCO	&	3860$-$7912	&	0.686	&3.7	        &	0.1	&	1.99	&	2.77	&	0.99	&	0.72 \\
03/25/20	&	8934.558	&	TCO	&	3860$-$7912	&	0.817	&4.9         &	0.2	&	1.81	&	2.65	&	0.96	&	0.68 \\
\noalign{\smallskip}\hline\noalign{\smallskip}
\end{tabular}
\end{center}
\begin{list}{}
\item Log of spectroscopic observations of 3\,Pup. Full Table is shown in the electronic version of the paper.
Column information: (1) -- Calendar date (MM/DD/YY), (2) -- Julian Date (JD$-$2450000), (3) - Observatory ID, (4) -- spectral range observed,
(5) -- orbital phase according to the RV solution (see text), (6) -- radial velocity in km\,s$^{-1}$ derived by cross-correlation in the range 4460--4632 \AA\, with respect to the template spectrum of 02/17/2017 (see Fig.\,\ref{f1}), (7) uncertainty in the radial velocity determination in km\,s$^{-1}$,
(8--11) -- parameters of the H$\alpha$ lines profiles: blue peak intensity in continuum units (I$_{\rm V}$), red peak intensity (I$_{\rm R}$), intensity of the central depression (I$_{\rm d}$), and the peak intensity ratio ($V/R$).
\item Comments on the spectra with no RV measurements: $^a$ -- a large portion or the entire cross-correlation region was not observed; $^b$ -- a systematic error in the wavelength calibration; $^c$ -- region damaged by reflection on the CCD chip.
\end{list}
}

\end{document}